\documentclass[10pt,aps,prd,a4paper,preprintnumbers,floatfix,nofootinbib,showpacs,superscriptaddress,notitlepage]{revtex4-1}
\usepackage[english]{babel}
\usepackage{amsmath}
\usepackage{graphicx}
\usepackage{color}
 \usepackage[table]{xcolor} 
\usepackage{mathrsfs}   
\usepackage{amssymb}
\usepackage{hyperref}
\usepackage[normalem]{ulem} 
\usepackage{dcolumn}
\usepackage{bm}
\usepackage{graphicx}
\usepackage[sort&compress]{natbib}
\usepackage{epsfig}
\usepackage{epstopdf}
\newcommand {\be} {\begin{equation}}
\newcommand {\ee} {\end{equation}}

 \def\snu{\tilde{\nu}}
 \newcommand{\beqa}{\begin{eqnarray}}
 \newcommand{\eeqa}{\end{eqnarray}}
 \newcommand{\ba}{\begin{array}} 
 \newcommand{\ea}{\end{array}}

\definecolor{greenLinks}{rgb}{0, 0.6, 0} 
\definecolor{blueLinks}{rgb}{0, 0, 0.6}
\definecolor{redLinks}{rgb}{0.6, 0, 0}
\definecolor{tempText}{rgb}{0.55, 0.10,0.67}
\definecolor{eprintLinks}{rgb}{0.4, 0.4, 0.4}
\definecolor{journalLinks}{rgb}{0.6, 0, 0}

 \def\lsim{\;\raise0.3ex\hbox{$<$\kern-0.75em\raise-1.1ex\hbox{$\sim$}}\;}
 \def\gsim{\;\raise0.3ex\hbox{$>$\kern-0.75em\raise-1.1ex\hbox{$\sim$}}\;}

\def\21{$\mathrm{SU(2)_L \otimes U(1)_Y}$ }
\def\31{$\mathrm{SU(3)_c \otimes U(1)_Q}$ }

\def\3211{$\mathrm{SU(3) \otimes SU(2)_L \otimes U(1)_R \otimes U(1)_{B-L}}$ }
\def\321{$\mathrm{SU(3) \otimes SU(2) \otimes U(1)}$ }
\def\422{$\mathrm{SU(4) \otimes SU(2) \otimes SU(2)_R}$ }
\newcommand {\ignore}[1]{}

\newcommand{\sm}{{Standard Model}}

\def\lfv{lepton flavour violation }
\def\cpv{CP violation }

\def\ss{\tilde{S}}

\begin{document} 
\vspace*{1cm}
\title{Inverse seesaw mechanism with compact supersymmetry:\\ enhanced naturalness and light super-partners}
\bigskip
\author{Valentina De Romeri}
\email{deromeri@ific.uv.es}
\affiliation{AHEP Group, Instituto de F\'{\i}sica Corpuscular, CSIC/Universitat de Val\`encia, Calle Catedr\'atico Jos\'e Beltr\'an, 2 E-46980 Paterna, Spain}
\author{Ketan M. Patel}
\email{ketan@iisermohali.ac.in}
\affiliation{Indian Institute of Science Education and Research Mohali, Knowledge City, Sector  81, S A S Nagar, Manauli 140306, India}

\author{Jose W. F. Valle}
\email{valle@ific.uv.es}
\affiliation{AHEP Group, Instituto de F\'{\i}sica Corpuscular, CSIC/Universitat de Val\`encia, Calle Catedr\'atico Jos\'e Beltr\'an, 2 E-46980 Paterna, Spain}

\begin{abstract}

We consider the supersymmetric inverse seesaw mechanism for neutrino mass generation within the context of a low energy effective theory where supersymmetry is broken geometrically in an extra  dimensional theory. It is shown that the effective scale characterizing the resulting compact supersymmetric spectrum can be as low as 500-600 GeV for moderate values of $\tan\beta$. The potentially large neutrino Yukawa couplings, naturally present in inverse seesaw schemes, enhance the Higgs mass and allow the super-partners to be lighter than in compact supersymmetry without neutrino masses. The inverse seesaw structure also implies a novel spectrum profile and couplings, in which the lightest supersymmetric particle can be an admixture of isodoublet and isosinglet sneutrinos. Dedicated collider as well as dark matter studies should take into account such specific features.

\end{abstract} 

\maketitle

\newpage
\section{Introduction}
\label{intro}

The discovery of a light Higgs boson \cite{Aad:2015zhl} has turned weak-scale supersymmetry (SUSY) into a leading candidate for a theory beyond the \sm~(SM). The minimal supersymmetric version of the standard model, also known as MSSM, is meant to provide a solution to the gauge hierarchy problem in addition to a radiative electroweak symmetry breaking mechanism with a light Higgs boson and a successful prediction for the weak mixing angle. However, the non-observation of any signal associated to weak-scale supersymmetry, particularly in the searches carried out by the ATLAS \cite{Aaboud:2017vwy,Aaboud:2017hrg,Aaboud:2017dmy,Aaboud:2018ujj} and CMS \cite{Sirunyan:2017pjw,Sirunyan:2017cwe,Sirunyan:2017kqq,Sirunyan:2018vjp} experiments at the Large Hadron Collider (LHC) so far, have raised concerns in the particle physics community at large.

We emphasize here that even if the existence of supersymmetry is confirmed in the on-going experiments, the MSSM does not provide a complete picture of physics beyond the SM, as two critical questions remain to be answered: what is an underlying mechanism of SUSY breaking? and what causes neutrinos to have mass?  Concerning the first question, it has been argued that the breaking of supersymmetry may arise from higher dimensional theories in which the extra spatial dimension(s) are compactified on orbifold~\cite{Scherk:1978ta,Scherk:1979zr}. Specific models of SUSY breaking based on this mechanism in five spacetime dimensions were constructed in \cite{Barbieri:2001yz}. Moreover, it has been recently shown that such models contain sufficient conditions which lead to relatively compressed SUSY spectrum if the compactification scale is not very far from the SUSY breaking scale \cite{Murayama:2012jh,Chowdhury:2016qnz}. A compact MSSM spectrum, in particular with an approximate degeneracy between the masses of gluino, squarks and neutralinos, is known to remain weakly constrained by direct searches \cite{LeCompte:2011cn,LeCompte:2011fh,Dreiner:2012sh,Dreiner:2012gx}. The MSSM with approximate degeneracy between all the sparticle masses, namely the Degenerate MSSM (DMSSM), has also been shown to satisfy various indirect constraints, allowing for a SUSY breaking scale as low as 700 GeV \cite{Chowdhury:2016qnz}.

Concerning the second issue, here we stress an important novel feature that emerges when neutrino masses arise from the seesaw mechanism~\cite{Minkowski:1977sc,GellMann:1980vs,Yanagida:1979as,mohapatra:1980ia,Schechter:1980gr,Schechter:1981cv,Lazarides:1980nt} realized at low-scale, such as the inverse and linear seesaw realizations~\cite{Mohapatra:1986bd,gonzalezgarcia:1989rw,Akhmedov:1995ip,Akhmedov:1995vm,Malinsky:2005bi}. It has been long ago noted that in such schemes \lfv and \cpv rates are unsuppressed by the small neutrino masses~\cite{Bernabeu:1987gr,Branco:1989bn}. The associated phenomenology has been widely discussed in the recent literature see, for example, Refs.~\cite{Deppisch:2004fa,Deppisch:2005zm,Abada:2011hm,Abada:2012cq,Deppisch:2013cya,Arganda:2014dta,Arganda:2015naa}.

In this paper, we consider the combined effects of having a compact supersymmetric spectrum as well as a low-scale seesaw origin of neutrino masses. For definiteness, we take the latter as the simplest inverse seesaw scheme in supersymmetry~\cite{Deppisch:2004fa}. While by itself the first mechanism allows for a lower effective SUSY scale, in conjunction with the second this effect is further enhanced, lowering the supersymmetric masses down to 500-600 GeV for moderate values of $\tan\beta$. The large Dirac-type Yukawa couplings allowed by the small neutrino masses within this setup enable a more natural way to account for the observed value of the Higgs boson mass 
\cite{Elsayed:2011de,Guo:2013sna,Chun:2014tfa} than in the simplest weak-scale degenerate SUSY scenario without neutrino masses.

Moreover, we note that in contrast to the standard MSSM case --- which does not accomodate neutrino masses --- the lightest  sneutrino state may also be a viable cold dark matter (DM) candidate, in addition to the widely studied case of the neutralino. When such a sneutrino, mainly made of ``right-handed'' or singlet components, is the lightest SUSY particle (LSP), it behaves as a weakly interacting massive particle (WIMP). Its interesting phenomenology has been widely studied in the literature~\cite{Arina:2007tm,Arina:2008bb,bazzocchi:2009kc,DeRomeri:2012qd,Chatterjee:2017nyx} also in the context of less minimal SUSY extensions, such as the NMSSM~\cite{Kitano:1999qb,Cerdeno:2008ep,Cerdeno:2017sks}.

This paper is organised as follows: section~\ref{sec:model} contains a description of the model, including details of the SUSY breaking sector and the mass spectra. In section~\ref{sec:num_an}, we discuss how our dedicated numerical analysis of the model is performed. In the same section we also list the experimental constraints applied on the model parameter space. In section~\ref{sec:spectra}, we thoroughly discuss the mass spectrum features resulting from our numerical analysis. Finally, in section~\ref{sec:DM}, we collect our results concerning the mixed sneutrino DM phenomenology. A final summary is given in section~\ref{sec:conclusions}.

\section{Model}
\label{sec:model}

In order to implement the inverse seesaw mechanism, the MSSM is extended by three generations of pairs of SM singlet superfields, $\hat\nu_i^c$ and $\hat S_i$. Under a global $U(1)$ symmetry corresponding to lepton number, the superfields $\hat\nu_i^c$ and $\hat S_i$ carry a charge $+1$ and $ -1$, respectively. The superpotential of the model can be written as
\be \label{W}
{\cal W} = {\cal W}_{\rm MSSM} + Y_\nu^{ij} \hat L_i\, \hat \nu_j^c\, \hat H_u + M_R^{ij}\, \hat \nu^c_i\, \hat S_j\, + \frac{1}{2} \mu_S^{ij}\, \hat S_i\, \hat S_j~, \ee
where the indices $i,j=1,2,3$ denote three generations. The $Y_\nu$, $M_R$ and $\mu_S$ are in general $3\times 3$ complex matrices, with $\mu_S$ symmetric. ${\cal W}_{\rm MSSM}$ denotes the standard MSSM superpotential:
\be\label{eq:WMSSM}
{\cal W}_{\rm MSSM} = Y_u^{ij}\,\hat{u}_i\,\hat{Q}_j\,\hat{H}_u\,
             + Y_d^{ij} \,\hat{d}_i\,\hat{Q}_j\,\hat{H}_d\,
             + Y_e^{ij} \,\hat{e}^c_i\,\hat L_j\, \hat{H}_d\,
             +\mu\,\hat{H}_u\,\hat{H}_d\ .
\ee

The above superpotential induces small neutrino masses in the following way. After the electroweak symmetry breaking, the $9 \times 9$ neutrino mass matrix can be written --- at tree level and in the $(\nu_i, \nu^c_i, S_i)$ basis --- as
\be \label{}
{\cal M}_{\nu} = \left( 
\begin{array}{ccc}
0 &m_D^{T}  &0\\ 
m_D  &0 &M_R\\ 
0 &M_{R}^{T} &\mu_S\end{array} 
\right) 
\ee
in which $m_D = \frac{1}{ \sqrt{2}} v_u Y_\nu$ is a Dirac mass term for the $\nu$, $\nu^c$ fields. Unlike  the case of type-I seesaw mechanism, the smallness of active neutrino masses here is attributed to the smallness of the elements of $\mu_S$ which characterize lepton number violation. This is in accordance with 't Hooft naturalness, since the limit $\mu_S \to 0$ restores the lepton number symmetry. The parameter $\mu_S$ can also be generated dynamically~\cite{bazzocchi:2009kc,DeRomeri:2017oxa,Chang:2017qgi}. The 9$\times$9 neutrino mass matrix can be diagonalised by a unitary matrix $U_\nu$, leading to 9 physical Majorana states. For $\mu_S\ll m_D\ll M_R$, the effective mass matrix $m_\nu$ of three light neutrinos is given by the following inverse seesaw relation \cite{Mohapatra:1986bd,gonzalezgarcia:1989rw}
\be
m_\nu = m_D^T {M_R^T}^{-1} \mu_S M_R^{-1} m_D\,.
\label{eq:effNumass}
\ee
The masses of the three heavy quasi-Dirac neutrino pairs are dominated by $M_{R}$ and they are given by  $\simeq M_R \pm \mu_S$. In short, the presence of non-vanishing $\mu_S$ introduces lepton number violation in the model, which give rise to small masses for the SM neutrinos, through the inverse seesaw mechanism \cite{Deppisch:2004fa}. The similar mechanism may also be realized in left-right symmetric extensions of the SM \cite{Akhmedov:1995ip,Akhmedov:1995vm,Malinsky:2005bi}.

\subsection{Supersymmetry breaking sector}
The most general soft breaking of supersymmetry can be parametrised by the following Lagrangian:
\beqa 
\label{soft}
-{\cal L}_{\rm soft} &=& -{\cal L}^{\rm MSSM}_{\rm soft} + (m^2_{\snu^c})_{ij}\, \snu^c_i \snu^c_j + (m^2_{\ss})_{ij}\, \ss_i \ss_j \nonumber \\
& + & A_\nu  Y_\nu^{ij} \tilde{L}_i\, \snu_j\, H_u + B_{M_R} M_R^{ij}\, \snu^c_i \ss_j\, + \frac{1}{2} B_{\mu_S} \mu_S^{ij}\, \ss_i \ss_j\,, 
\eeqa
where ${\cal L}^{\rm MSSM}_{\rm soft}$ contains the generic soft supersymmetry breaking terms in the MSSM. A compact SUSY spectrum at the weak scale can be obtained if the soft masses of gauginos, squarks and sleptons are taken to be approximately equal \cite{Chowdhury:2016qnz}. The soft masses are also required to be real and flavour universal in order to comply with non-observation of any statistically significant evidence of flavour or CP violation other than those predicted by the SM. These conditions are naturally realized in models of SUSY breaking based on the Scherk-Schwarz mechanism \cite{Scherk:1979zr}. In this case, one begins with $N=1$ supersymmetry in five dimensional spacetime where the extra spatial dimension is compactified on a circle of radius $R$. An orbifold is then constructed by introducing a $Z_2$ symmetry under which the extra dimensional coordinate transforms as $ y \to - y$. This gives rise to two fixed points: $y=0$ and $y= \pi R$. The $N=1$ SUSY in 5D is equivalent to an effective $N=2$ supersymmetry in 4D \cite{ArkaniHamed:2001tb}. The $Z_2$ symmetry of the orbifold is used to break one of these two supersymmetries \cite{Barbieri:2001yz}. The remaining 4D, $N=1$ supersymmetry is broken by the so-called twist under which the superpartners of the SM fields are assumed to be non-cyclic, for example $\phi(y+2\pi R) =e^{ 2 \pi i \alpha} \phi(y)$, where $\alpha$ ($0 \le \alpha < 1$) is a twist parameter.  A non-vanishing value of $\alpha$  generates massive modes of these fields on the fixed points. The SM fields are assumed to be cyclic (with $\alpha = 0$) which result into their massless modes at the fixed points.

If the matter and gauge fields are localized in the bulk and Higgs fields are introduced at a fixed point, the above way of SUSY breaking results into the following soft masses at the compatification scale $1/R$  \cite{Barbieri:2001yz}:
\be \label{ss1}
M_1 = M_2 = M_3
=\frac{\alpha}{R} \equiv M_S \ee
\be \label{ss2}
m^2_{\tilde{q}}=m^2_{\tilde{u^c}}=m^2_{\tilde{d^c}}=
m^2_{\tilde{l}}=m^2_{\tilde{e^c}}=m^2_{\snu^c} = m^2_{\ss}=M_S^2\, {\mathbb{I}} \ee
\be \label{ss3}
A_0=A_\nu = -2 M_S\,,~~B_{M_R}=B_{\mu_S}=-2 M_S^2, \ee
\be \label{ss4}
m_{H_u}^2=m_{H_d}^2=0\,,~~B_\mu=0\,. 
\ee
Here, we use the conventional MSSM notation in which $M_i$ are gaugino masses, $m_{\tilde f}$ is a $3 \times 3$ mass matrix of sfermion of kind $\tilde{f}$ and $A_0$ is universal trilinear scalar coupling. The above universality in the soft masses leads to approximate degeneracy in the physical mass spectrum of supersymmetric particles \cite{Chowdhury:2016qnz}. It is important to note that the running effects in the soft masses, from the mediation scale $1/R$ to the SUSY breaking scale $M_S=\alpha/R$, can introduce large non-degeneracy in the soft masses and therefore one requires $\alpha \approx 1$ in order to obtain a compact SUSY spectrum. 

The vanishing value of $m_{H_u}^2$, $m_{H_d}^2$ and $B_\mu$ is due to the fact that the Higgs superfields are localized on a brane and therefore they do not feel the effect of supersymmetry breaking at the leading order. Non-zero values of these parameters are required to trigger electroweak symmetry breaking. This can be achieved by either considering radiative corrections at the scale $1/R$ \cite{Murayama:2012jh} or taking into account the presence of brane localised source of soft SUSY breaking. The latter choice essentially makes $m_{H_u}^2$, $m_{H_d}^2$ and $B_\mu$ free parameters, avoiding the constraint given in Eq. (\ref{ss4}). 

The parameters in the superpotential do not get fixed by the SUSY breaking mechanism and they do not have any dependence on $\alpha/R$ in general. However in practice, this leads to a non-degeneracy in the sparticle spectrum. For example in the MSSM, for $\mu \ll M_S$ one gets some of the neutralinos/charginos much lighter than the common SUSY scale $\sim M_S$. This leads to a large mass gap between the gluino (or stop) and the lightest neutralino. To avoid this problem we make the phenomenologically viable choice $\mu \approx \alpha/R$, as it is advocated in \cite{Chowdhury:2016qnz}.

\subsection{Physical mass spectrum of sparticles}
\label{sparticle_spectrum}

The physical mass spectrum of SUSY particles which arises from the soft masses given in Eqs. (\ref{ss1}-\ref{ss3}) is discussed in \cite{Chowdhury:2016qnz}. The masses of the first and second generations of squarks and charged sleptons are almost degenerate in this case. Because of the presence of large trilinear terms, the masses of the third generation sfermions get modified significantly in comparison to those of the first two generations. For example, large $A_t \equiv A_0 y_t$ induces large mixing among the stops and their masses get split by $m_{\tilde{t}_2}^2-m_{\tilde{t}_1}^2 \approx 2 m_t |A_t - \mu \cot\beta|$ \cite{Chowdhury:2016qnz}. The large mixing among the top squarks also helps in obtaining relatively higher Higgs mass. 

With the conditions in Eq. (\ref{ss1}) and $M_S \gg M_Z$, one obtains two of the neutralinos  with masses $\sim M_S$ and the remaining with mass $\sim |\mu|$. As discussed earlier, the choice $\mu \approx M_S$ leads to an approximate degeneracy between all the neutralinos. The same choice also implies degeneracy between charginos. The splitting in the masses of neutralinos and charginos is induced by the contributions from electroweak symmetry breaking. Therefore, the departure from degeneracy becomes significant if $M_S$ is close to the electroweak scale. For $M_S \gg M_Z$, all the gauginos have almost degenerate masses of ${\cal O}(M_S)$. 

In comparison to the Degenerate MSSM presented earlier in \cite{Chowdhury:2016qnz}, one of the distinct features in this model is the presence of sneutrinos. There are 18 sneutrino mass eigenstates. Depending on their masses one of them can be the LSP. This would be stable due to $R$-parity conservation, and therefore it could be the candidate for cold DM. It is convenient to separate the sneutrino mass matrix into CP-even and CP-odd blocks \cite{Grossman:1997is}\footnote{The difference between the eigenvalues of the real and imaginary components of the sneutrinos is a lepton number violating mass term \cite{Hirsch:1997vz}, analogous to the $\mu_S$ term.} such that
\begin{eqnarray}
{\cal M}^2  =
\left(\begin{array}{cc} 
{\cal M}^2_+ & {\bf 0} \\
 {\bf 0} & {\cal M}^2_-
\end{array}\right),
\label{eq:mSnupn}
\end{eqnarray}
in the CP eigenstates basis $(\tilde{\nu}^*_+, \tilde{\nu}^{c*}_+, S^*_+, \tilde{\nu}^*_-, \tilde{\nu}^{c*}_-, S^*_-)$. At the tree-level, the mass matrices for the scalar neutrinos ${\cal M}_\pm^2$ are \cite{Arina:2008bb,Hirsch:2009ra,DeRomeri:2012qd}
\begin{eqnarray}
{\cal M}_{\pm}^2  = 
\left(\begin{array}{ccc} 
m^2_{\tilde{l}}+ D^2 + (m_D^Tm_D)  & (A_\nu -\mu\cot\beta) m_D^T & m_D^T M_R \\
(A_\nu -\mu\cot\beta) m_D & m^2_{\tilde{\nu}^c}+(M_RM_R^{T})+(m_Dm_D^{T}) & 
       \pm M_R\mu_S + B_{M_R} \\
M_R^T m_D & \pm \mu_S M_R^{T} + B_{M_R}^T 
& m^2_{\tilde S}+ \mu_S^2+M_R^TM_R \pm B_{\mu_S}
\end{array}\right)
\label{eq:mSnu}
\end{eqnarray}
where  $m^2_{\tilde{l}}$, $m^2_{\tilde{\nu}^c}$ and $m_{\tilde S}^2$ are scalar soft masses and $D^2=\frac{1}{2} m^2_Z \cos 2\beta$. Despite of degeneracy in soft masses, the above matrix can lead to non-degeneracy in sneutrino masses, depending on the values of right handed neutrino masses. For example, for a single generation of $(\tilde{\nu}^*_\pm, \tilde{\nu}^{c*}_\pm, S^*_\pm)$ and using the conditions given in Eq. (\ref{ss2},\ref{ss3}) together with $\mu_S \ll M_Z, M_D \ll M_R, M_S$, one finds that the determinant of the $3 \times 3$ matrix ${\cal M}_{\pm}^2$ given above is approximated as 
\begin{eqnarray} 
\label{analytical_sneutrino}
{\rm Det}.[{{\cal M}_{+}^2}] & \approx & M_S^6 \left(-5 + \frac{M_R^4}{M_S^4} \left(1+ {\cal O}\left(\frac{m_D^2}{M_S^2}\right) +  {\cal O}\left(\frac{D^2}{M_S^2} \right)+ ...  \right) + ... \right)\,, \nonumber \\
{\rm Det}.[{{\cal M}_{-}^2}] & \approx & M_S^6 \left(-1 + \frac{4 M_R^2}{M_S^2} \left(1+ {\cal O}\left(\frac{m_D^2}{M_S^2}\right) +  {\cal O}\left(\frac{D^2}{M_S^2} \right)+ ...  \right) + ... \right)\,.
\end{eqnarray} 
One obtains a relatively small value of ${\rm Det}.[{\cal M}_{+}^2]$ or  ${\rm Det}.[{\cal M}_{-}^2]$ for $M_S \approx 0.67 M_R$ or $M_S \approx 2 M_R$, respectively. In these cases, the cancellation within the terms in Eq. (\ref{analytical_sneutrino}) leads to a very light sneutrino with mass well below the degenerate scale $M_S$. Such a light sneutrino is the LSP and remains stable because of R-parity conservation. Therefore, this scenario gives rise to a novel possibility in which a relatively light sneutrino (with $m_{\tilde{\nu}_{\rm LSP}} \lesssim 100$ GeV) can be a viable DM candidate, while the remaining SUSY spectrum is approximately degenerate. This is quite different from the compact SUSY frameworks discussed previously in \cite{Chowdhury:2016qnz,Murayama:2012jh}, in which the neutralino is the DM and the spectrum degeneracy enforces its mass to be $\sim M_S$.

\section{Numerical analysis}
\label{sec:num_an}
In order to study the physical mass spectrum and the effects of various direct and indirect searches on the allowed parameters of the model we now perform dedicated numerical analyses. Most soft masses, trilinear and bilinear parameters follow the degeneracy conditions given in Eqs.~(\ref{ss1},\ref{ss2},\ref{ss3}) imposed by the Scherk-Schwarz mechanism.  As mentioned earlier, the parameters $m_{H_{u,d}}^2$ and $B_\mu$ remain undetermined by the mechanism if brane localized SUSY breaking terms are introduced. We choose 
$$ \mu = M_S\,,~~B_\mu = 2 M_S^2$$
and determine the values of $m_{H_{u,d}}^2$ by solving the tadpole equations leading to consistent radiative electroweak symmetry breaking. We assume that such values of $m_{H_{u,d}}^2$ and $B_\mu$ parameters are generated by introducing an adequate SUSY breaking sector on the fixed point on  orbifold. The above choice of $\mu$ parameter implies approximate degeneracy in the masses of charginos and neutralinos, as discussed in the previous section. We also fix the sign of $\mu$ and $B_\mu$ parameters, assuming them to be positive.

For the parameters in the neutrino sector, it is convenient to use the parametrization introduced in \cite{Arganda:2014dta}, in which the $3\times 3$ matrix $\mu_S$ is fixed by inverting the seesaw relation: 
\begin{equation}
\label{MuSparametrization}
\mu_S=M_R^T m_D^{-1} U^* m_\nu  U^\dagger m_D^{T^{-1}} M_R,
\end{equation}
where $m_\nu={\rm Diag.}(m_{\nu_1},m_{\nu_2},m_{\nu_3})$ are the masses of the three light neutrinos and $U$ is the lepton mixing matrix characterizing neutrino oscillations, assumed unitary as an approximation. The above parametrization allows to choose $M_R$ and $Y_\nu$ as input parameters. One can choose diagonal and real $M_R$ without loss of generality. Once the values of parameters in $Y_\nu$ and $M_R$ are fixed, $\mu_S$ is determined using the global fit values of neutrino masses and mixing angles from \cite{deSalas:2017kay}. 

The various considerations and assumptions made above leave the following as free parameters in the model,
$$ M_S,\, \tan\beta,\, M_{R_1},\, M_{R_2},\, M_{R_3},\, Y_\nu^{ij},$$
where $M_{R_i}$ are ``right-handed'', mainly singlet, neutrino masses. We will consider four different benchmark scenarios with particular choices for the values of $M_{R_i}$ and $Y_{\nu}^{ij}$. These are listed in Table \ref{tab:benchmarks_in}. 
\begin{table}[h!]
\begin{center}
\footnotesize{
{\begin{tabular}{| c | c | c | c | c |}  \hline    
Benchmark & $M_S$ [GeV] & tan$\beta$ &  $M_{R_i}$ [GeV] & $Y_\nu$ \\
 \hline   \hline 
P1 & [400 - 1300] & [5 - 30] & (1000, 1200, 1400) & diag(0.5, 0.6, 0.3) \\
P2 & [400 - 1300] & [5 - 30] & (2000, 2200, 2400) & diag(0.5, 0.6, 0.3) \\
P3 & [400 - 1300] & [5 - 30] & (1000, 1200, 1400) & diag(0.7, 0.8, 0.5) \\
P4 & [400 - 1300] & [5 - 30] & (2000, 2200, 2400) & diag(0.7, 0.8, 0.5) \\
\hline                       
\end{tabular}}
}
\caption{Input parameters for different benchmark points.}
\label{tab:benchmarks_in}
\end{center}
\end{table}
We assume diagonal and real $Y_\nu$ with couplings of ${\cal O}(1)$ and consider, as examples, two sets of values for such couplings. For each of these choices, two example right-handed neutrino mass spectra are considered. For each case, we vary $M_S$ in the range from 400 to 1300 GeV and $\tan\beta$ in the range $5$ to $30$, as displayed in Table \ref{tab:benchmarks_in}.  

The above framework is first implemented in  {\tt SARAH 4.9.1}~\cite{Staub:2013tta}. We then calculate the physical particle spectrum with {\tt SPheno 4.0.3}~\cite{Porod:2003um,Porod:2011nf}. We further use {\tt Micromegas 5.0.2} \cite{Belanger:2014vza} to compute the thermal component to the sneutrino DM relic abundance. Note that the input parameters are defined at the compactification scale, which is assumed to be close to the SUSY breaking scale in order to avoid splitting from renormalization group evolution. Therefore, we neglect ``running'' in the soft parameters. {\tt SARAH  4.9.1}  calculates all vertices, mass matrices, tadpoles equations, one-loop corrections for tadpoles and self-energies for the model. {\tt SPheno} calculates the SUSY spectrum using low energy data and the supplied model as input. Notice that the masses are calculated at 2-loops.  The flavour observables are computed with the {\tt FlavorKit} \cite{Porod:2014xia} extension of {\tt SARAH}.\\[-.2cm]

After estimating the physical mass spectrum and various observables, we take into account the following relevant constraints from various direct and indirect searches. \\[-.2cm]

\paragraph{Neutrino oscillation data}:
We require compatibility of our inverse seesaw model with the best-fit intervals for the neutrino oscillation parameters. This is implemented through Eq. (\ref{MuSparametrization}) in which the masses in $m_\nu$ and mixing parameters in $U$ are taken from the results of the recent global fit to the neutrino oscillation data given in \cite{deSalas:2017kay}. The yet undetermined Dirac and Majorana phases in $U$ are set to zero for definiteness. We assume a normal ordering of light neutrino masses with the lightest active neutrino mass $m_{\nu_1} = 0.01$ eV. 

\paragraph{Direct searches}:
So far the data from the LHC have not shown any indication of supersymmetric particles in direct searches. These data provide the strongest constraints on the masses of colored superpartners, namely squarks and gluino. However, these bounds are typically obtained in simplified schemes in which several specific assumptions are made for the masses of chargino and neutralinos and different branching ratios. For example, the latest analyses from ATLAS \cite{Aaboud:2017vwy,Aaboud:2017hrg,Aaboud:2017dmy,Aaboud:2018ujj} made using $36.1$ fb$^{-1}$ data collected at $\sqrt{s} = 13$ TeV disfavour gluino (squarks) with masses upto $1.85$ ($1.3$) TeV. These hold in simplified schemes if the neutralino is massless. Similar studies from $35.8$ fb$^{-1}$ data collected at $\sqrt{s} = 13$ TeV by the CMS collaboration \cite{Sirunyan:2017pjw,Sirunyan:2017cwe,Sirunyan:2017kqq,Sirunyan:2018vjp} yield lower bounds on the masses of gluino $\sim 2$ TeV, third generation squarks $\sim 1$ TeV and first two generation squarks $\sim 1.3$ TeV again, for massless neutralino in simplified schemes. However, most of these stringent constraints  become much weaker if the SUSY spectrum is compressed \cite{LeCompte:2011cn,LeCompte:2011fh,Dreiner:2012sh,Dreiner:2012gx}. If the mass difference between squarks/gluino and the neutralino lies within $100$-$200$ GeV, the lower bounds on the masses of squarks and gluino can be significantly lower. For example, in such cases the masses of third (first two) generation squarks can be as small as $450$ ($600$) GeV, while the gluino can be as light as $750$ GeV, as inferred by the latest ATLAS \cite{Aaboud:2018ujj} and CMS \cite{Sirunyan:2018vjp} analyses. We do not impose any such direct constraints on our spectrum, as these analyses assume specific decay channels as well as branching ratios which could be quite different in the specific model under consideration. Instead, we explicitly give the complete spectrum for the lowest value of $M_S$ allowed by the other constraints listed below for each of our benchmark scenarios. The resulting benchmark spectrum is found to be consistent with the limits discussed above.

\paragraph{Higgs boson mass}: One of the most important constraints on the parameter space of our model comes from the LHC measurement of the Higgs mass \cite{Aad:2015zhl}. The lightest CP-even Higgs boson in the MSSM is identified with the discovered scalar particle with mass close to 126 GeV.  It is well-known that the observed value of Higgs mass requires significant contributions from higher loops involving SUSY particles, see for example \cite{Hall:2011aa,Heinemeyer:2011aa,AlbornozVasquez:2011aa,Arbey:2011aa,Arbey:2011ab,Draper:2011aa,Carena:2011aa}. In particular, such contributions require either multi-TeV squarks or large trilinear coupling in the top sector. The latter is naturally arranged in our framework as seen from Eq. (\ref{ss3}). Furthermore, the presence of neutrinos in the inverse seesaw model provides additional contribution to the Higgs mass at one-loop \cite{Elsayed:2011de,Chun:2014tfa}. This effect helps in reproducing the $126$ GeV Higgs mass in our framework with relatively lighter stops. In our numerical analysis, the Higgs mass is computed using {\tt SPheno 4.0.3} which includes full two-loop calculation using a diagrammatic approach with vanishing external momenta~\cite{Goodsell:2015ira}. In order to take into account the theoretical uncertainty in the estimation of Higgs mass, we allow $\pm 3$ GeV deviation from its experimentally measured value when comparing it to the model's prediction.

\paragraph{Invisible Higgs decay width}:  The properties of the Higgs boson observed at the LHC have been shown to be consistent with the predictions of the SM. Additional contributions to the Higgs boson width from non-SM decay channels can be constrained with the branching fraction of the Higgs boson decaying into lighter stable particles that interact very weakly with the detector~\cite{Shrock:1982kd,Joshipura:1992hp,DeCampos:1994fi,Bonilla:2015jdf,Bonilla:2015uwa}. In particular, the possibility that the lightest neutral Higgs boson in this model can decay invisibly into a pair of mixed sneutrino LSPs (with mass $\lesssim m_{h^0}/2$) places an important constraint~\cite{Banerjee:2013fga}. Current limits from ATLAS and CMS on the  invisible branching fraction of the Higgs boson (assuming a SM Higgs boson production cross section) are around BR$(h^0 \to \rm inv) \lesssim 20\% - 30\%$~\cite{Tanabashi:2018}.

\paragraph{Flavour observables in the B sector}:
Indirect constraints from flavour physics experiments, such as B factories and LHCb 
are often sensitive to high SUSY mass scales. For instance, the decay of a strange $B$ meson ($ B_s$) into two oppositely charged muons is very rare in the SM. Hence, this branching fraction is sensitive to new physics such as SUSY. The observation of $B_s \to \mu^+ \mu^-$ from the combined analysis of CMS and LHCb data is  BR($B_s \rightarrow \mu^+ \mu^-) = (2.7^{+0.6}_{-0.5}) \times 10^{-9}$ \cite{Patrignani:2016xqp}.  We can define $ R_{b_s\mu\mu} = \frac{{\rm BR}(B_s \to \mu^+ \mu^-)}{{\rm BR}(B_s \rightarrow \mu^+ \mu^-)_{\rm SM}}$, where the SM prediction is BR$(B_s \to \mu^+ \mu^-)_{\rm SM} = (3.65 \pm 0.23) \times 10^{-9}$~\cite{Bobeth:2013uxa}. Throughout our analysis we apply $R_{b_s\mu\mu} = 0.74 \pm 0.17$ considering its $3\sigma$ range. The measurement of the branching fraction of $B \to X_s \gamma$ is currently performed with quite a good accuracy: BR$(B \to X_s \gamma) = (3.32 \pm 0.16) \times 10^{-4}$~\cite{Amhis:2016xyh}. Using the SM prediction at the next-to-next-to-leading-order prediction ${\rm BR}(B \to X_s \gamma)_{\rm SM} = (3.36 \pm 0.23) \times 10^{-4}$, we get an allowed range for the ratio $R_{bs\gamma} = \frac{{\rm BR}(B \to X_s \gamma)}{{\rm BR}(B \to X_s \gamma)_{\rm SM}} = 0.99 \pm 0.08$. We also apply this constraint at 3$\sigma$.

\paragraph{Lepton flavour violating observables}:
The non-observation of flavour violations in the charged lepton sector can also be used in order to restrict new physics models. In particular, rare decays and transitions such as the decay of the muon have been widely discussed within inverse seesaw models, with and without supersymmetry~\cite{Deppisch:2004fa,Deppisch:2005zm,Abada:2011hm,Abada:2012cq,Deppisch:2013cya,Arganda:2014dta,Abada:2014nwa,Abada:2014cca,Abada:2014kba,Abada:2015oba,Abada:2015zea,Arganda:2015naa,DeRomeri:2016gum}. We apply the most stringent limit to date on the branching fraction of this rare muon decay, which has been set by the MEG experiment: BR$(\mu \to e \gamma) < 4.2 \times 10^{-13}$~\cite{TheMEG:2016wtm}.

\paragraph{Dark matter}:
We adopt a standard cosmological scenario, where the mixed sneutrino DM particles were in thermal equilibrium with the SM ones in the early Universe. Hence, if the mixed sneutrino is the only DM particle contributing to the cosmological DM, its relic density must fall within the cosmological range for cold DM derived by the Planck analysis~\cite{Ade:2015xua,Aghanim:2018eyx}: $ 0.117 \leq \Omega_{\tilde{\nu}_{\rm LSP}} \leq  0.123$ (3$\sigma$ range). If other DM candidates are simultaneously present together with the mixed sneutrino then its relic abundance should be $  \Omega_{\tilde{\nu}_{\rm LSP}} < 0.117$. We note that our mixed sneutrino DM scenario can be probed in direct detection (DD) experiments, which are designed to detect the nuclear recoil in the scattering of galactic sneutrinos off target nuclei (see for instance \cite{Goodman:1984dc}). The signal rate depends on astrophysical quantities (subject to considerable uncertainties), such as the local density and velocity distribution of sneutrinos in our galaxy and, from the particle physics side, on the sneutrino LSP mass and on the scattering cross section \cite{Arina:2008bb,DeRomeri:2012qd}. The current most stringent limit on WIMP-nucleon spin-independent (SI) elastic scattering cross section has been set with 278.8 days of data collected by the XENON1T experiment at LNGS \cite{Aprile:2018dbl}.\\

Apart from the above restrictions, there are several other indirect constraints on new physics, for instance from hadronic and leptonic flavour physics data and the anomalous magnetic moment of the muon $(g-2)_\mu$. We however do not include all of them since those discussed in the previous paragraphs are known to have dominant effects on the MSSM with low scale SUSY. The list of constraints imposed on the parameter space of our model are summarised in Table \ref{tab:constraints}.
\begin{table}[h!]
\begin{center}
\begin{tabular}{| c | c | c | c | c | c | c | c | c | c |}  \hline 
Observable & Constraint applied\\
\hline \hline
$m_{h^0}$ & $123 \leq m_{h^0} \leq 129$ GeV\\       
$R_{bs\gamma}$ &  $0.75 <R_{bs\gamma} < 1.23$\\
$R_{B_s\mu \mu}$ &$0.23 <R_{B_s\mu \mu} < 1.25$ \\
BR($\mu \to e \gamma$) &  $< 4.2\cdot10^{-13}$ \\
$\Omega_{\tilde{\nu}_{\rm LSP}} h^2$ & $\leq 0.123$ \\ 
 \hline                       
\end{tabular}
\caption{Main experimental constraints applied on the model parameter space.}
\label{tab:constraints}
\end{center}
\end{table}

The results obtained from the numerical analysis, as discussed in the previous sections, are outlined in the next sections. We first discuss the particle spectra in various benchmark scenarios and then we provide a detailed discussion of the mixed sneutrino DM phenomenology.

\section{Particle spectra} 
\label{sec:spectra}
\begin{figure}[!ht]
\begin{center}
\begin{tabular}{cc}
\epsfig{file=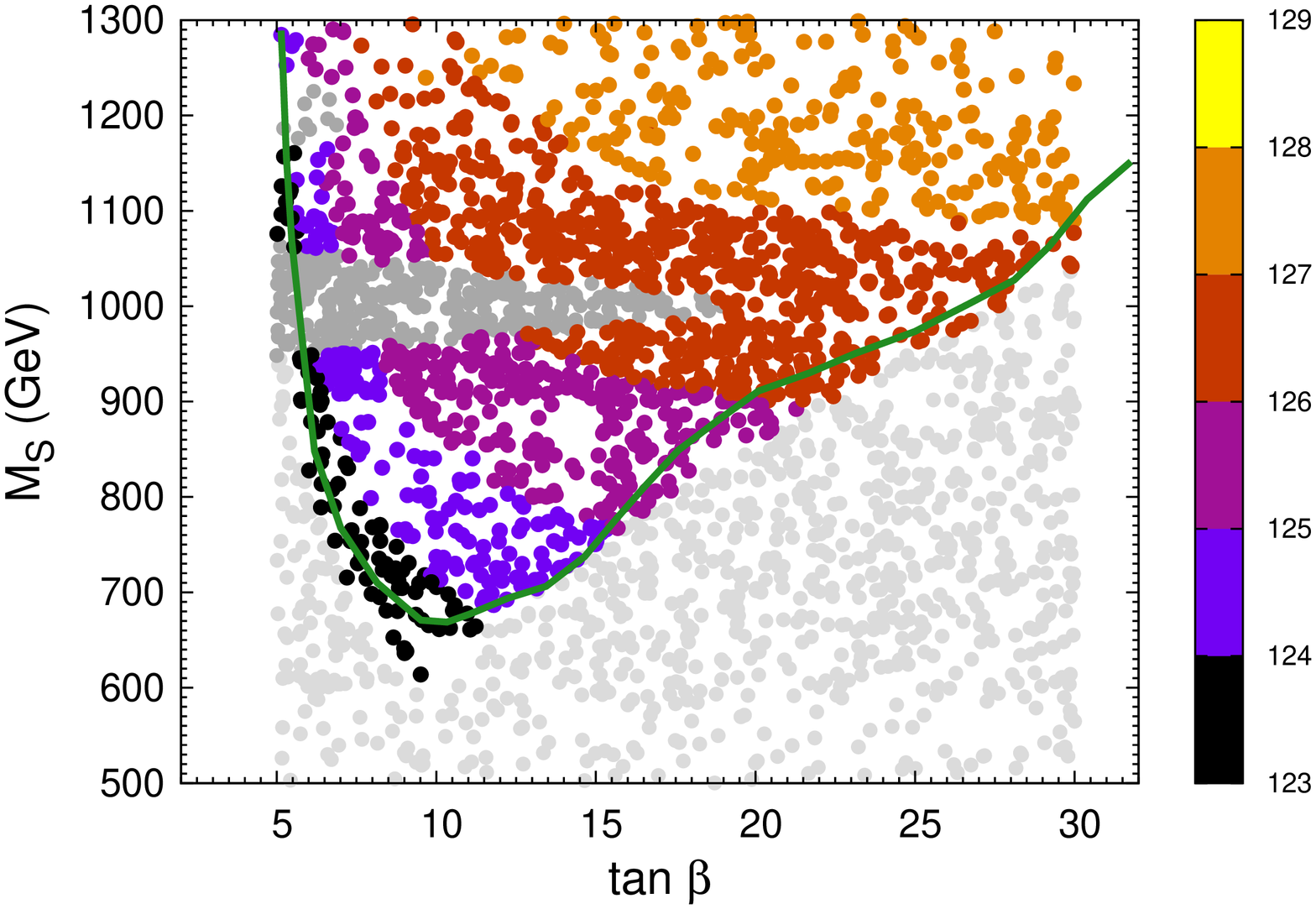,
width=53mm, angle =270} 
\hspace*{2mm}&\hspace*{2mm}
\epsfig{file=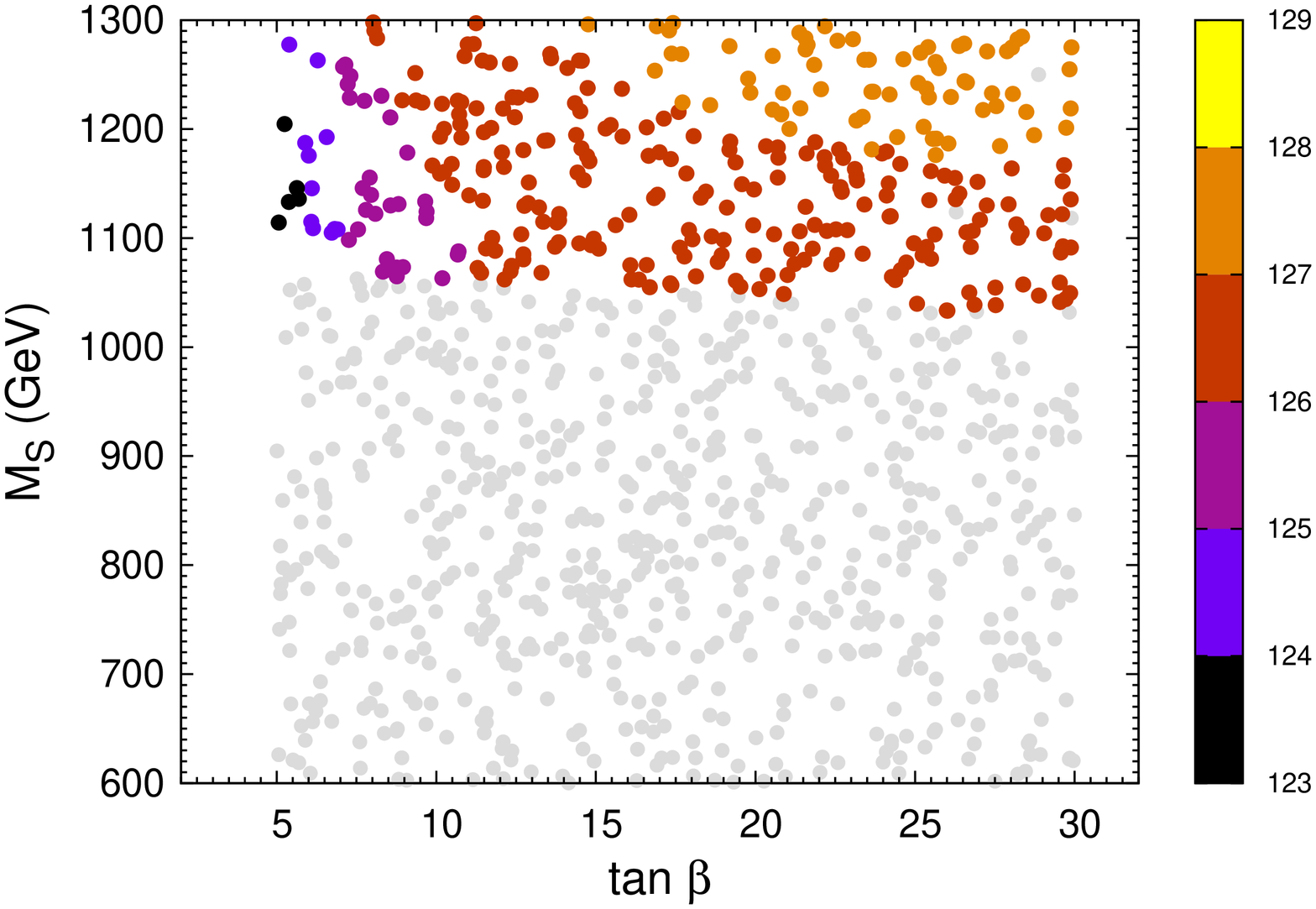,
width=53mm, angle =270}
\\
\epsfig{file=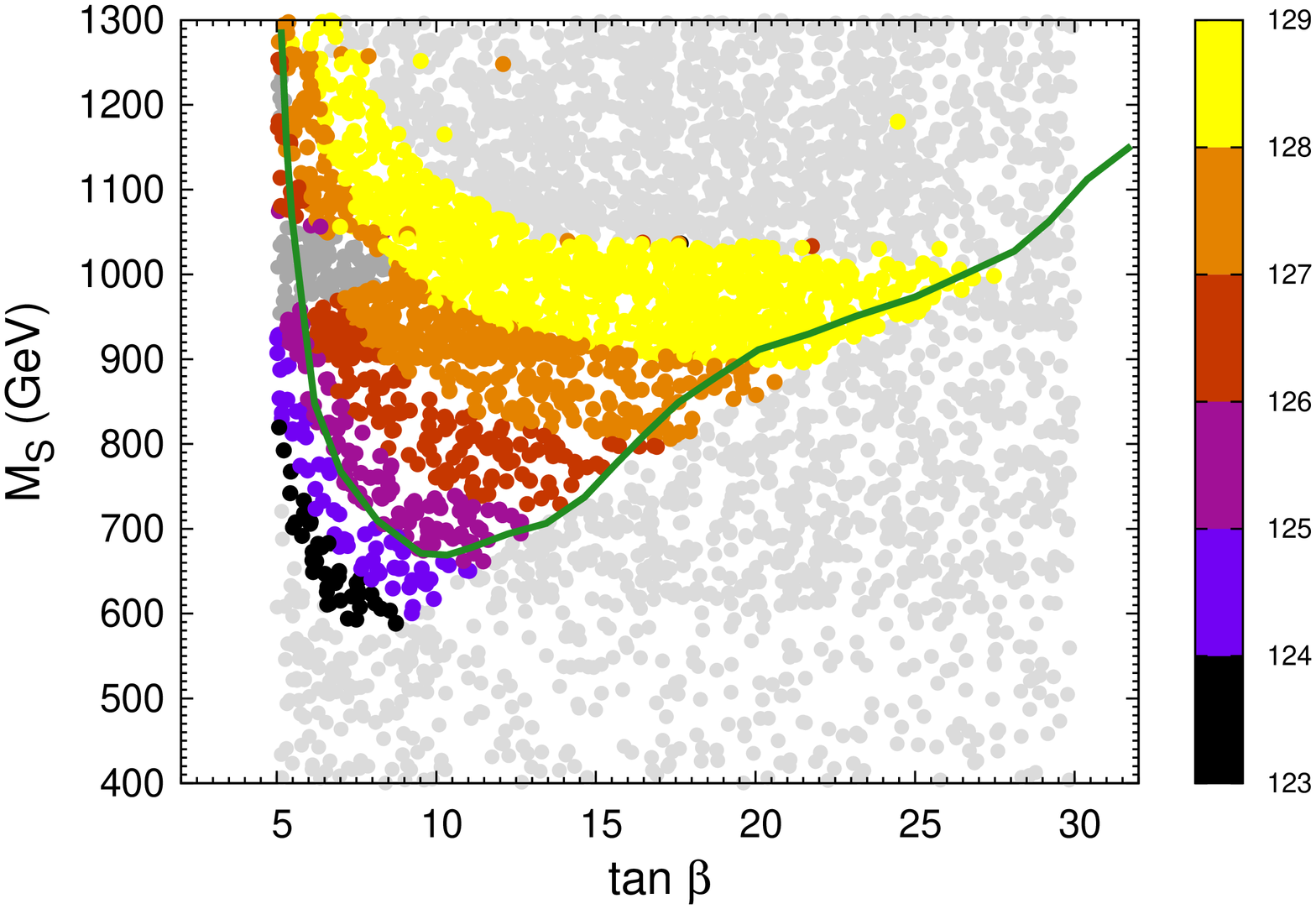,
width=53mm, angle =270}
\hspace*{2mm}&\hspace*{2mm}
\epsfig{file=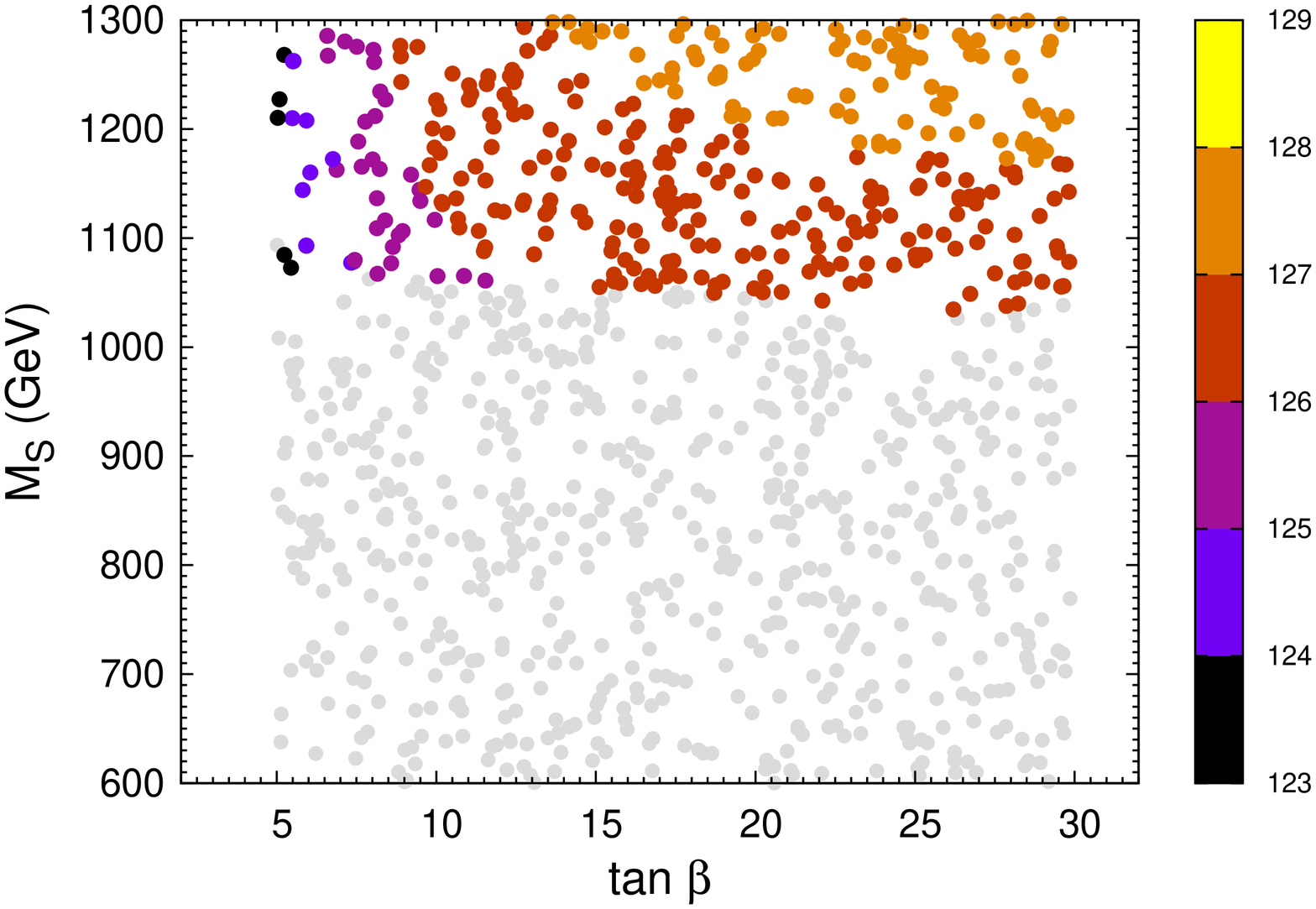,
width=53mm, angle =270}
\end{tabular}
\end{center}
\caption{Predicted Higgs boson mass in the plane tan$\beta$ - $M_S$. The colour scheme denotes regimes for the $m_{h^0}$ in GeV. Grey points are excluded by at least one of the constraints listed in Table \ref{tab:constraints}. Dark grey points refer to mass spectra with $m_{\tilde{\nu}_{\rm LSP}}< m_{h^0}/2$. (We do not impose the constraint on the sneutrino relic density.) From the top-left to the down-right corner, the four panels refer to the benchmark scenarios P1 to P4 respectively, see Table~\ref{tab:benchmarks_in}. The green contour in P1 and P3 corresponds to the lower limit on $M_S$ obtained from the Higgs mass constraints with $Y_\nu = 0$ (i.e. without inverse seesaw mechanism).}
\label{fig:mH_alpha_tb}
\end{figure}

As discussed earlier, one of the main features of this model is an enhancement in the Higgs mass provided by the presence of neutrinos with Yukawa couplings of ${\cal O}(1)$. In Fig. \ref{fig:mH_alpha_tb}, we display the constraints on $\tan\beta$ and $M_S$ arising from the Higgs mass for the four different benchmark scenarios defined in Table \ref{tab:benchmarks_in}.  
In all the panels, the grey points correspond to solutions excluded by at least one of the constraints listed in Table \ref{tab:constraints}.  Dark grey points denote solutions which in general lead to BR($h^0 \to $ inv) $\gtrsim 0.3$. (We do not impose the constraint on the sneutrino relic density, which we will study in more detail in the next section.) It can be seen that the Higgs mass increases with $M_S$ and it stays close to its experimentally measured value for $800 \lesssim M_S \lesssim 1300$ GeV and $7 \lesssim \rm tan \beta \lesssim 20$ (P1), $1050 \lesssim M_S \lesssim 1300$ GeV and $7 \lesssim \rm tan \beta \lesssim 12$ (P2), $650 \lesssim M_S \lesssim 1000$ GeV and $5 \lesssim \rm tan \beta \lesssim 13$ (P3), $1050 \lesssim M_S \lesssim 1300$ GeV and $5 \lesssim \rm tan \beta \lesssim 8$ (P4). The lowest value of the SUSY-breaking scale --- in agreement with the constraint on the Higgs mass --- is $M_S \sim 700$ GeV in P1 and $M_S \sim 600$ GeV in P3. The Higgs mass increases for larger values of $M_S$ due to an enhancement in the loop contributions from relatively heavier stops.

In the panels corresponding to the benchmark scenarios P1 and P3, a (green) contour corresponding to the lower limit on $M_S$ obtained from the Higgs mass constraints with $Y_\nu = 0$ is displayed, in order to compare our results with those of the standard massless neutrino case given in \cite{Chowdhury:2016qnz}. Clearly, the nonzero value of $Y_\nu$ helps in accommodating the observed Higgs boson mass with relatively lower $M_S$, as seen from the left panels in Fig. \ref{fig:mH_alpha_tb}. Notice that one obtains considerable enhancement in $m_{h^0}$ for P3 because of the larger $Y_\nu$ values assumed in this case with respect to those in P1 (see above). The extra contribution to the Higgs mass due to the presence 
of ${\cal O}(1)$ Dirac Yukawa couplings possible in the inverse seesaw model, allows slightly lower value of $M_S$ in comparison to the standard massless neutrino case. We have checked that, solutions with $M_S$ as low as $\sim 400$ GeV are possible, with $Y_\nu \sim 1$.  Such points are nevertheless subject to strong constraints from the direct searches at the LHC as described in section \ref{sec:num_an}.

For the excluded regions in panels P1 and P3, lowest values of $\tan\beta$ and $M_S$ are disfavoured by $m_{h^0} < 123$ GeV, while the bottom-right region corresponding to large $\tan\beta$ is excluded by the $B \to X_s \gamma$ constraint. The top-right region in P3 leads to $m_{h^0} > 129$ GeV and hence it is excluded. In the benchmark scenarios P2 and P4, sneutrinos are degenerate with the rest of the SUSY spectrum. This happens because in P2 and P4 the chosen range of $M_S$ and the values of the right-handed neutrino masses in $M_{R}$ do not satisfy either of the cancellation conditions discussed below Eq. (\ref{analytical_sneutrino}). Therefore, one does not get light sneutrino in this case. Instead we observe that for $M_S \lesssim 1$ TeV, usually stau or stop is the LSP and hence this region is disfavoured because it does not provide a viable DM candidate.

\begin{figure}[h!]
\begin{center}
\begin{tabular}{cc}
\epsfig{file=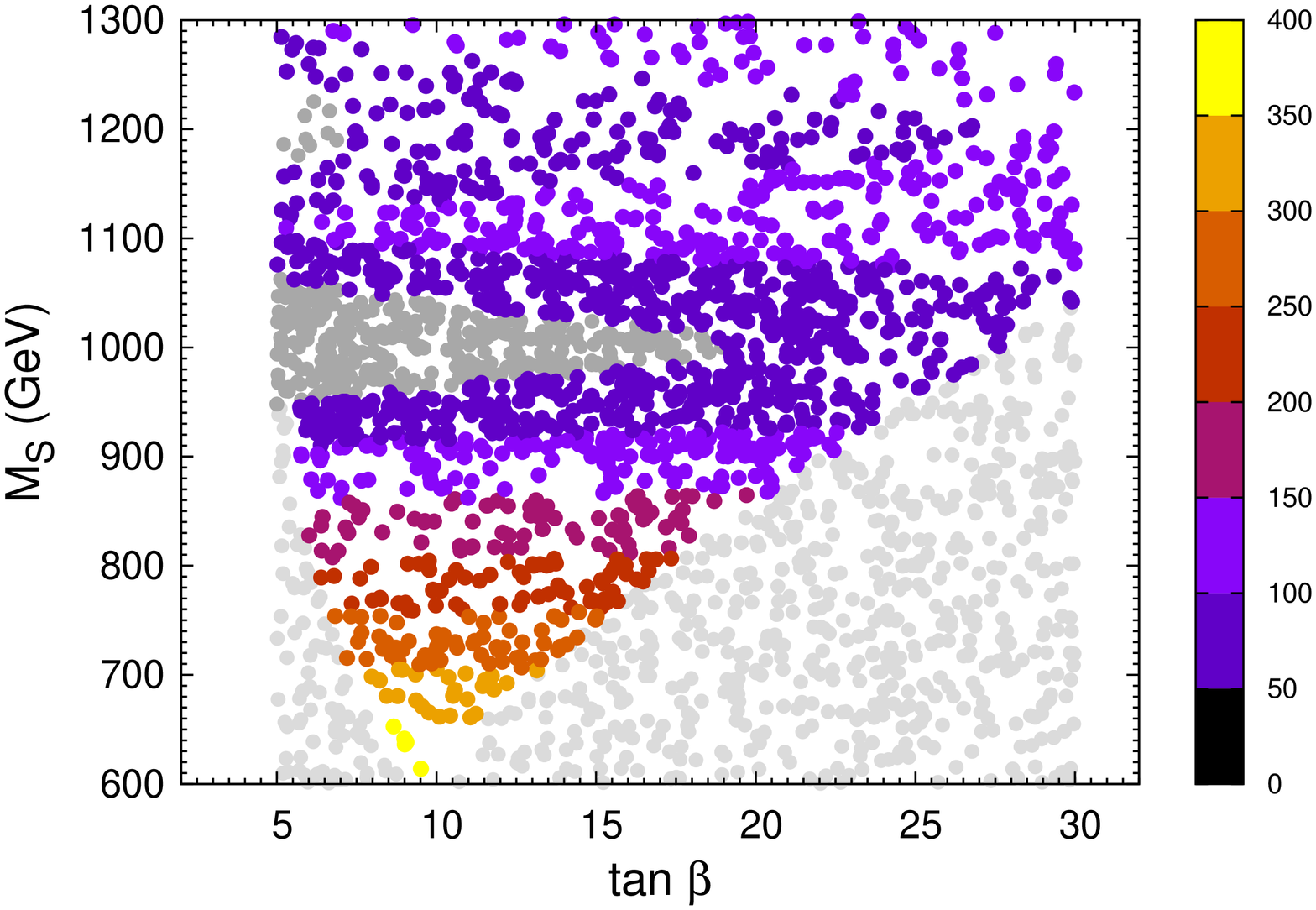,
width=53mm, angle =270} 
\hspace*{2mm}&\hspace*{2mm}
\epsfig{file=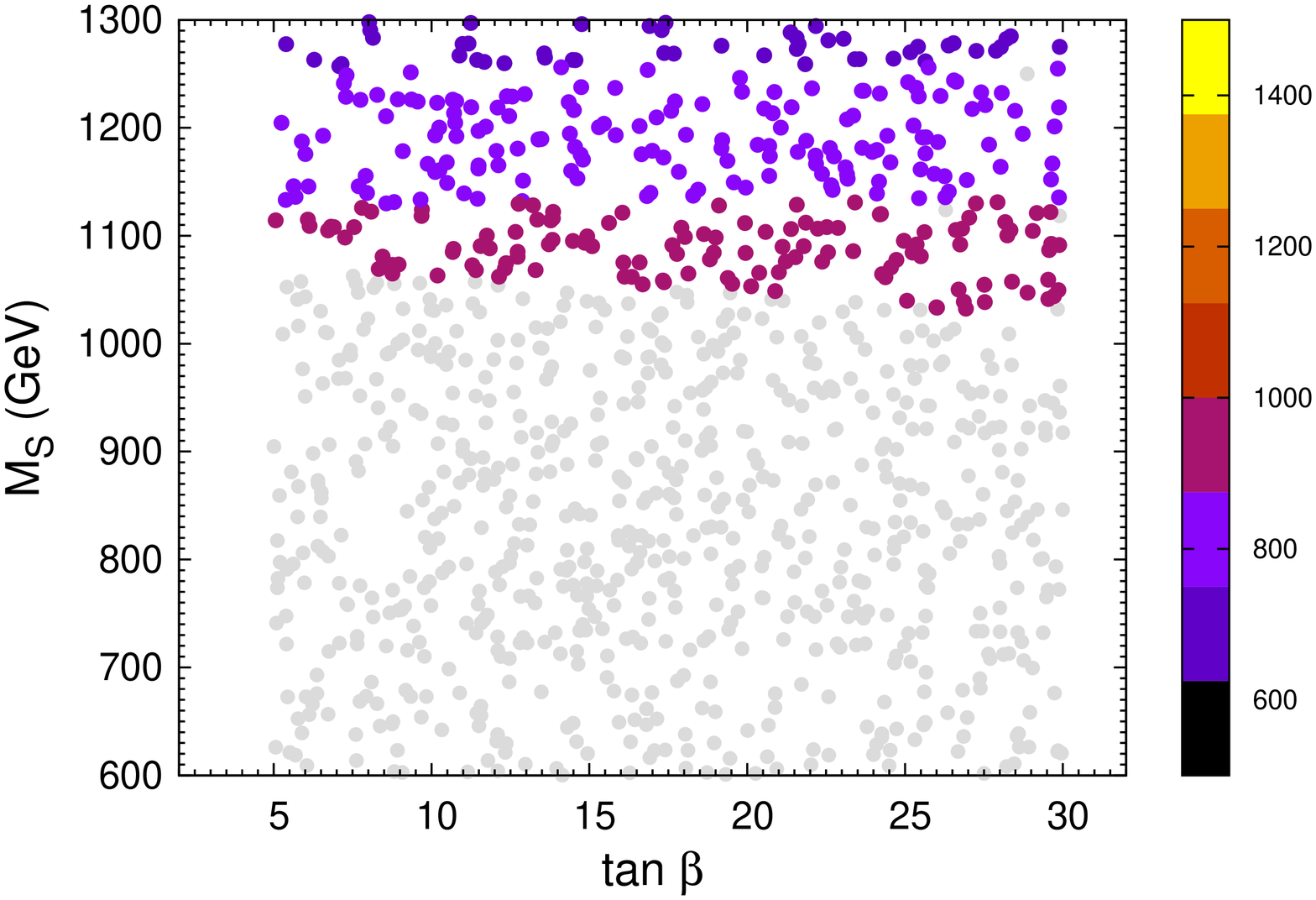,
width=53mm, angle =270}
\\
\epsfig{file=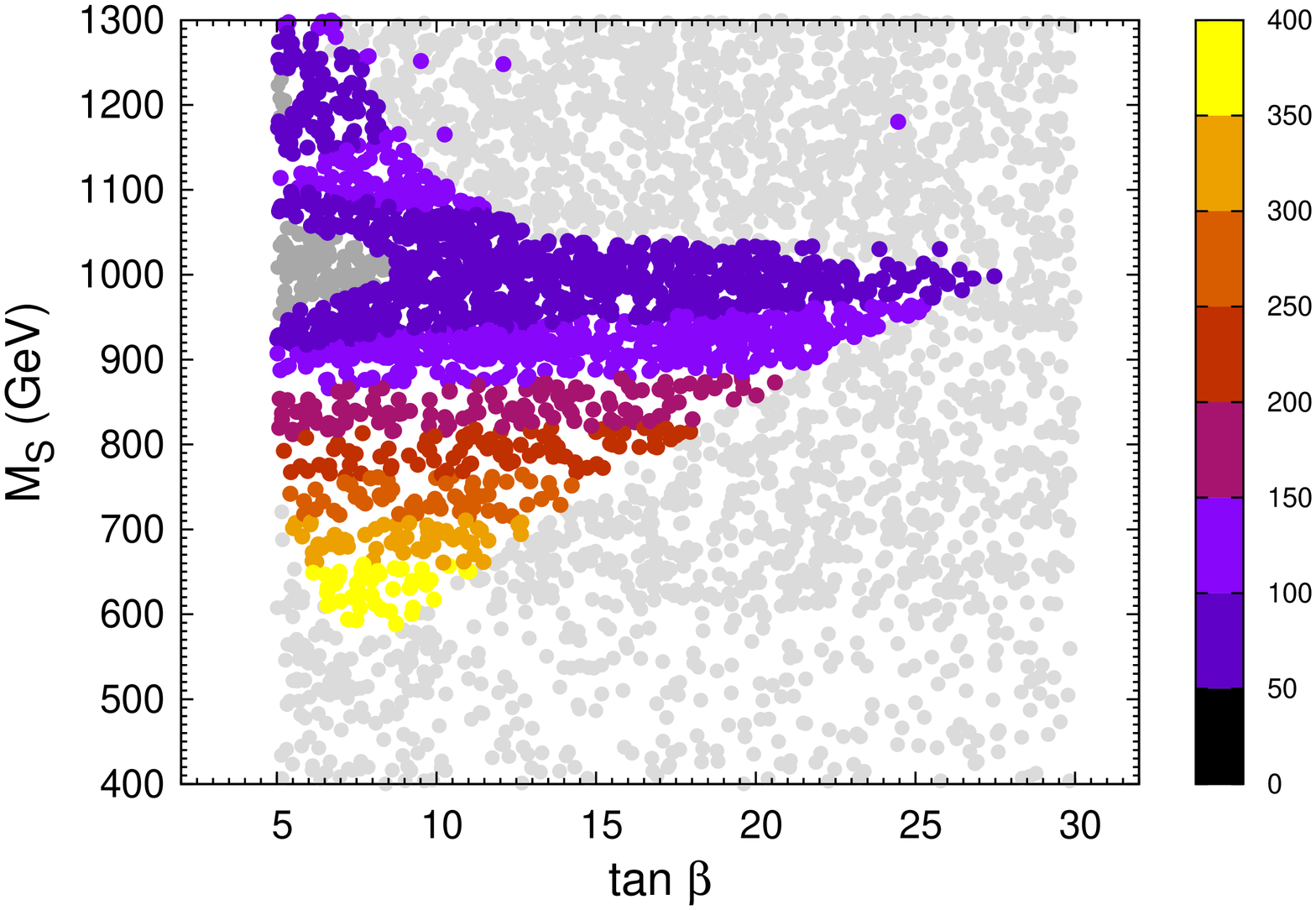,
width=53mm, angle =270}
\hspace*{2mm}&\hspace*{2mm}
\epsfig{file=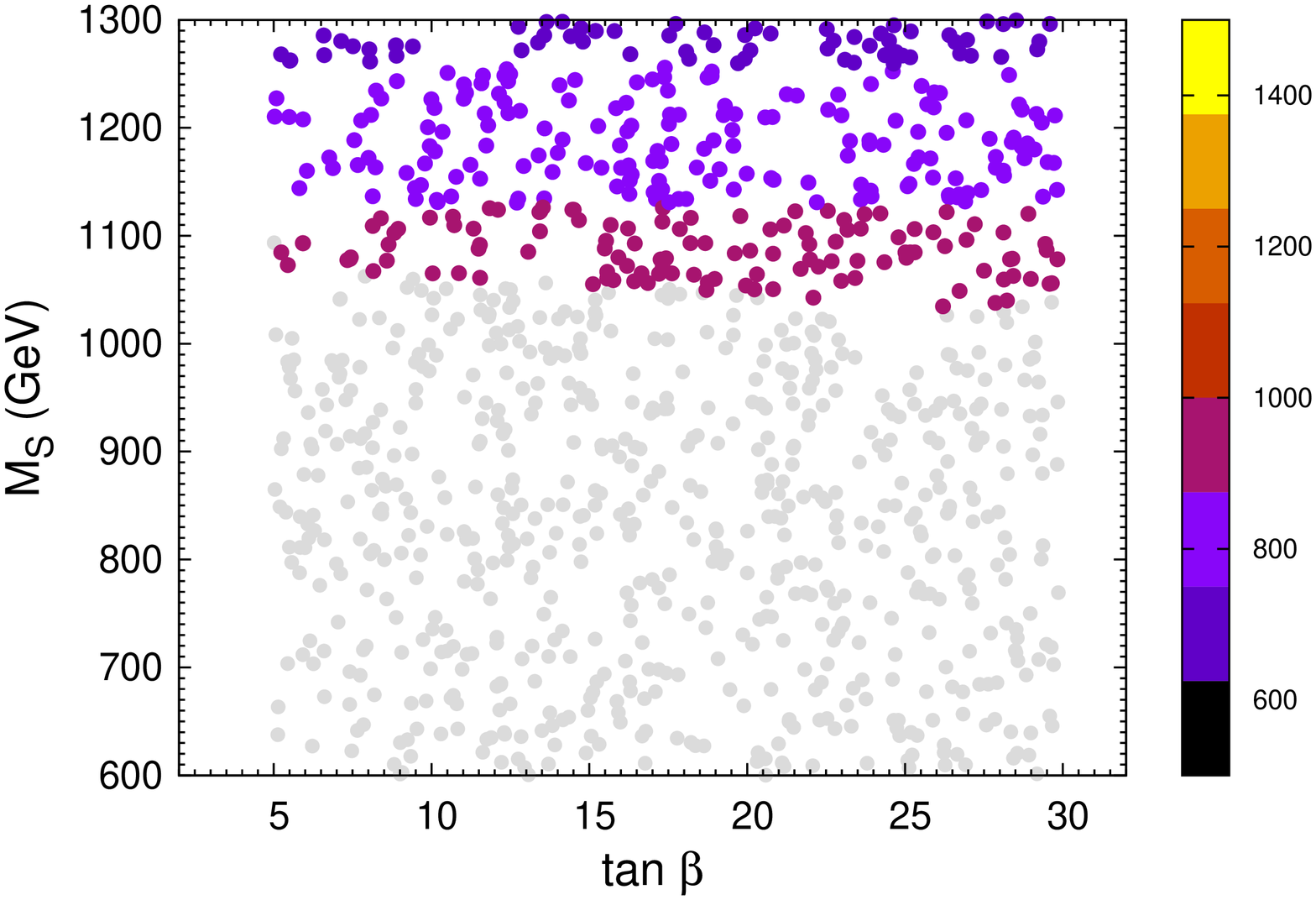,
width=53mm, angle =270}
\end{tabular}
\end{center}
\caption{Predictions for the sneutrino mass in the plane tan$\beta$ - $M_S$. The colour scheme denotes regimes for the $m_{\tilde{\nu}_{\rm LSP}}$ in GeV. Grey points denote exclusions due to at least one of the constraints listed in Table \ref{tab:constraints}. Dark grey points refer to mass spectra with $m_{\tilde{\nu}_{\rm LSP}}< m_{h^0}/2$. (We do not impose the constraint on the sneutrino relic density.) From top-left to bottom-right, the four panels refer to the benchmark scenarios P1 to P4 defined in Table~\ref{tab:benchmarks_in}.}.
\label{fig:msnu_alpha_tb}
\end{figure}

The masses of sneutrinos primarily depend on the parameters: $M_S$, $M_{R_i}$, $Y_\nu^{ij}$ and $\mu_S$. The values chosen for these parameters lead to most sneutrino masses lying in the TeV scale regime. However, as discussed in section \ref{sparticle_spectrum}, particular choices of $M_{R_i}$ and $M_S$ can give rise to very small masses for at least one of the sneutrinos, hence being the LSP. In Fig.~\ref{fig:msnu_alpha_tb}, we show the values of the sneutrino LSP mass, in the parameter space  tan$\beta$ - $M_S$.
\hspace*{-8mm}

Notice that the sneutrino LSP can be as light as few GeV in some regions of the parameter space. As discussed previously, the large off-diagonal entries in the sneutrino mass matrix of Eq.~(\ref{eq:mSnu}) lead to a large splitting between the eigenstates when $M_S \sim 0.67 M_R$. As can be seen from the choice of input parameters given in Table \ref{tab:benchmarks_in}, such a condition can be realized only in case of P1 or P3. This leads to the lightest sneutrino mass as small as $\sim 50$ GeV in these cases. In contrast, the ``cancellation'' does not happen for the benchmark scenarios P2 and P4. In this case the lightest sneutrino remains relatively heavier, as can be seen from Fig. \ref{fig:msnu_alpha_tb}. To summarise the discussion regarding the SUSY spectra found in our numerical scans, we present a complete spectrum for one representative point from each of the benchmark scenarios in Table~\ref{tab:spectraP1}. The most relevant observables are also listed in the same table.

\begin{table}[h!]
\begin{center}
{\begin{tabular}{| c | c | c | c | c | c | c | c | c | c |}  \hline                     
Particle & Mass [GeV] \\
 \hline   \hline 
$\tilde{d}_1$ & $709.8$ \\
$\tilde{d}_{2,3}$ & $720.3$ \\
$\tilde{d}_{4,5}$ & $729.5$ \\
$\tilde{d}_6$ & $740.7$ \\
$\tilde{u}_1$ & $565.4$ \\
$\tilde{u}_{2,3}$ & $725.2$ \\
$\tilde{u}_{4,5}$ & $733.6$ \\
$\tilde{u}_6$ & $870.1$ \\
$\tilde{l}_1$ & $683.2$ \\
$\tilde{l}_{2,3}$ & $684.5$ \\
$\tilde{l}_4$ & $713.8$ \\
$\tilde{l}_5$ & $717.6$ \\
$\tilde{l}_6$ & $717.9$ \\
$\tilde{\nu}^R_1$, $\tilde{\nu}^I_1$ & 311.3 \\
$\tilde{g}$ & $743.5$ \\
$h^0$ & 124.1 \\
$H, A^0$ & $3.33\cdot10^{3}$ \\
$\tilde{\chi}^0_1$ & $625.8$ \\
$\tilde{\chi}^+_1$ & $637.4$ \\
\hline \hline
Parameter & Value \\
 \hline   \hline 
 $M_S$ & 695 GeV\\
tan$\beta$ & 11.5 \\
\hline \hline
Observable & Value \\
 \hline   \hline 
$\Omega_{\tilde{\nu}_{\rm LSP}} h^2$ & $0.312$ \\
$(g-2)_\mu$ & $4.03 \cdot10^{-10}$ \\
$R_{bs\gamma}$ &  $0.77$\\
BR($\mu \to e \gamma$) &  $6.8\cdot10^{-35}$ \\
 $R_{B_s\mu \mu}$ & 0.98 \\
 \hline                       
\end{tabular}
}
\quad
{\begin{tabular}{| c | c | c | c | c | c | c | c | c | c |}  \hline                     
Particle & Mass [GeV] \\
 \hline   \hline 
$\tilde{d}_{1,2}$ & $1087.2$ \\
$\tilde{d}_{3}$ & $1094.5$ \\
$\tilde{d}_{4,5}$ & $1103.2$ \\
$\tilde{d}_6$ & $1140$ \\
$\tilde{u}_1$ & $958.8$ \\
$\tilde{u}_{2,3}$ & $1100.5$ \\
$\tilde{u}_{4,5}$ & $1125.1$ \\
$\tilde{u}_6$ & $1247$ \\
$\tilde{l}_{1,2}$ & $1022.6$ \\
$\tilde{l}_{3}$ & $1038.9$ \\
$\tilde{l}_4$ & $1095.9$ \\
$\tilde{l}_5$ & $1101.8$ \\
$\tilde{l}_6$ & $1104.6$ \\
$\tilde{\nu}^R_1$, $\tilde{\nu}^I_1$ & 947.8 \\
$\tilde{g}$ & $1129.4$ \\
$h^0$ & 126.4 \\
$H, A^0$ & $6.19\cdot10^{3}$ \\
$\tilde{\chi}^0_1$ & $991.6$ \\
$\tilde{\chi}^+_1$ & $1003.8$ \\
\hline \hline
Parameter & Value \\
 \hline   \hline 
 $M_S$ & 1056.8 GeV\\
tan$\beta$ & 17.4 \\
 \hline   \hline 
Observable & Value \\
 \hline   \hline 
$\Omega_{\tilde{\nu}_{\rm LSP}} h^2$ & $0.109$ \\
$(g-2)_\mu$ & $2.43 \cdot10^{-10}$ \\
$R_{bs\gamma}$ &  $0.85$\\
BR($\mu \to e \gamma$) &  $2.4\cdot10^{-35}$ \\
 $R_{B_s\mu \mu}$ & 0.99 \\
 \hline                       
\end{tabular}
}
\quad
{\begin{tabular}{| c | c | c | c | c | c | c | c | c | c |}  \hline                     
Particle & Mass [GeV] \\
 \hline   \hline 
$\tilde{d}_{1}$ & $689.5$ \\
$\tilde{d}_{2,3}$ & $700.3$ \\
$\tilde{d}_{4,5}$ & $707.6$ \\
$\tilde{d}_6$ & $708.7$ \\
$\tilde{u}_1$ & $525.8$ \\
$\tilde{u}_{2,3}$ & $703.4$ \\
$\tilde{u}_{4,5}$ & $705$ \\
$\tilde{u}_6$ & $847.5$ \\
$\tilde{l}_{1}$ & $669$ \\
$\tilde{l}_{2,3}$ & $671.2$ \\
$\tilde{l}_4$ & $691.2$ \\
$\tilde{l}_5$ & $691.7$ \\
$\tilde{l}_6$ & $697$ \\
$\tilde{\nu}^R_1$, $\tilde{\nu}^I_1$ & 335.9 \\
$\tilde{g}$ & $719.7$ \\
$h^0$ & 123.4 \\
$H, A^0$ & $2.36\cdot10^{3}$ \\
$\tilde{\chi}^0_1$ & $598.3$ \\
$\tilde{\chi}^+_1$ & $610.2$ \\
\hline \hline
Parameter & Value \\
 \hline   \hline 
 $M_S$ & 672.9 GeV\\
tan$\beta$ & 6.12 \\
 \hline   \hline 
Observable & Value \\
 \hline   \hline 
$\Omega_{\tilde{\nu}_{\rm LSP}} h^2$ & $0.109$ \\
$(g-2)_\mu$ & $2.66 \cdot10^{-10}$ \\
$R_{bs\gamma}$ &  $0.87$\\
BR($\mu \to e \gamma$) &  $1\cdot10^{-33}$ \\
 $R_{B_s\mu \mu}$ & 0.96 \\
 \hline                       
\end{tabular}
}
\quad
{\begin{tabular}{| c | c | c | c | c | c | c | c | c | c |}  \hline                     
Particle & Mass [GeV] \\
 \hline   \hline 
$\tilde{d}_{1,2}$ & $1092.7$ \\
$\tilde{d}_{3}$ & $1093$ \\
$\tilde{d}_{4,5}$ & $1107.8$ \\
$\tilde{d}_6$ & $1133.6$ \\
$\tilde{u}_1$ & $956.4$ \\
$\tilde{u}_{2,3}$ & $1105$ \\
$\tilde{u}_{4,5}$ & $1125.3$ \\
$\tilde{u}_6$ & $1246.7$ \\
$\tilde{l}_{1,2}$ & $1031.5$ \\
$\tilde{l}_{3}$ & $1041.9$ \\
$\tilde{l}_4$ & $1096.8$ \\
$\tilde{l}_5$ & $1102$ \\
$\tilde{l}_6$ & $1102.6$ \\ 
$\tilde{\nu}^R_1$, $\tilde{\nu}^I_1$ & 944.2 \\
$\tilde{g}$ & $1133$ \\
$h^0$ & 126.3 \\
$H, A^0$ & $5.86\cdot10^{3}$ \\
$\tilde{\chi}^0_1$ & $994.6$ \\
$\tilde{\chi}^+_1$ & $1006.7$ \\
\hline \hline
Parameter & Value \\
 \hline   \hline 
 $M_S$ & 1060 GeV\\
tan$\beta$ & 15.6 \\
 \hline   \hline 
Observable & Value \\
 \hline   \hline 
$\Omega_{\tilde{\nu}_{\rm LSP}} h^2$ & $0.111$ \\
$(g-2)_\mu$ & $2.16 \cdot10^{-10}$ \\
$R_{bs\gamma}$ &  $0.87$\\
BR($\mu \to e \gamma$) &  $1.3\cdot10^{-35}$ \\
 $R_{B_s\mu \mu}$ & 0.99 \\
 \hline                       
\end{tabular}
}
\caption{Particle spectra and relevant observables for four benchmark points from scenarios P1, P2, P3 and P4.}
\label{tab:spectraP1}
\end{center}
\end{table}

\section{Mixed sneutrino dark matter}
\label{sec:DM}

The novelty of embedding the inverse seesaw mechanism within the DMSSM is that the SUSY spectrum allows now for a bosonic DM candidate, which is the lightest sneutrino (in addition to the neutralino, which may also be allowed). We now proceed to discuss the phenomenology of the mixed sneutrino as DM candidate.

\paragraph{Relic density}

The relic abundance of the mixed sneutrino DM is a direct consequence of the strength of its annihilations. Thus, in full generality, the relic density of mixed sneutrinos depends on the magnitude of the Yukawa coupling $Y_\nu$. Indeed, should the sneutrinos be pure singlets ($\tilde{\nu}^{c}$ or $S$), they would not couple to gauge bosons and thus they would tend to be overabundant. For mixed sneutrinos, their relic density will depend on $Y_\nu$ and in general  $Y_\nu \gsim 0.1$ is required in order to produce sufficient annihilations so as to lower the relic density below the current cosmological bounds (unless different mechanisms, such as coannihilations are present)~\cite{DeRomeri:2012qd}. Nevertheless, $Y_\nu$ is also relevant in determining other observables, such as the Higgs mass and the radiative \lfv decays such as $\mu \to e \gamma$. Of course, in the latter case the flavor structure of $Y_\nu$ plays a key r\^ole.\\[-.2cm]

\begin{figure}[!htb]
\begin{center}
\epsfig{file=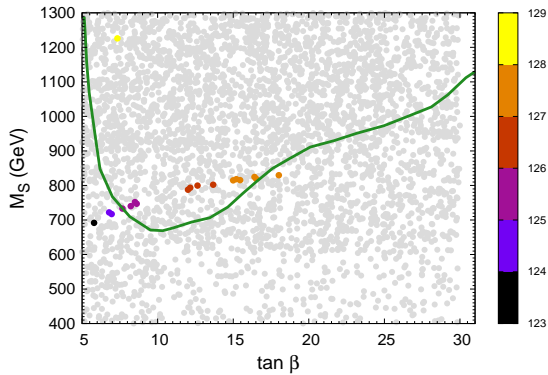,
width=53mm, angle =270} 
\end{center}
\caption{Predicted Higgs boson mass in the plane tan$\beta$ - $M_S$, for the benchmark scenario P3. The colour scheme denotes regimes for the $m_{h^0}$ in GeV. Only solutions in agreement with the dark matter relic density constraint (i.e. $0.117 < \Omega_{\tilde{\nu}_{\rm LSP}} h^2 < 0.123$) are shown as palette-coloured points. Moreover, grey points are excluded by at least one of the constraints listed in Table \ref{tab:constraints}. The green contour is the same as in Fig.~\ref{fig:mH_alpha_tb}.}
\label{fig:mH_omega_tanb_mS}
\end{figure}

We show in Fig.~\ref{fig:mH_omega_tanb_mS} the predicted Higgs boson mass in the plane tan$\beta$ - $M_S$, for the benchmark scenario P3. To illustrate the importance of the neutrino Dirac Yukawa coupling $Y_\nu$ both at enhancing
the Higgs boson mass and driving the relic abundance of the
sneutrino LSP, we keep only the solutions with $0.117 < \Omega_{\tilde{\nu}_{\rm LSP}} h^2 < 0.123$. The green contour is the same as in Fig.~\ref{fig:mH_alpha_tb}.

\hspace*{-8mm}

\begin{figure}[!htb]
\begin{center}
\begin{tabular}{cc}
\epsfig{file=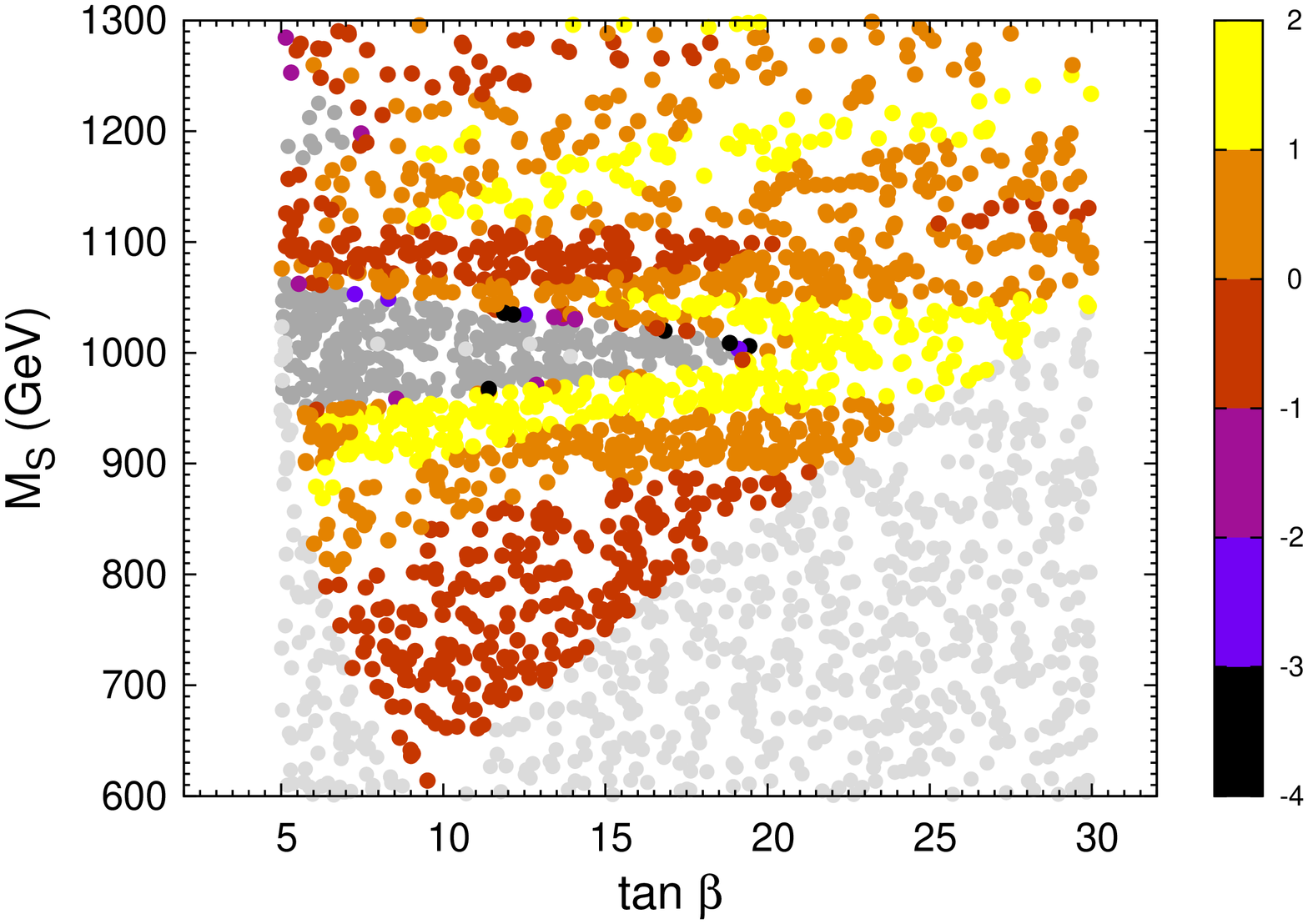,
width=53mm, angle =270} 
\hspace*{2mm}&\hspace*{2mm}
\epsfig{file=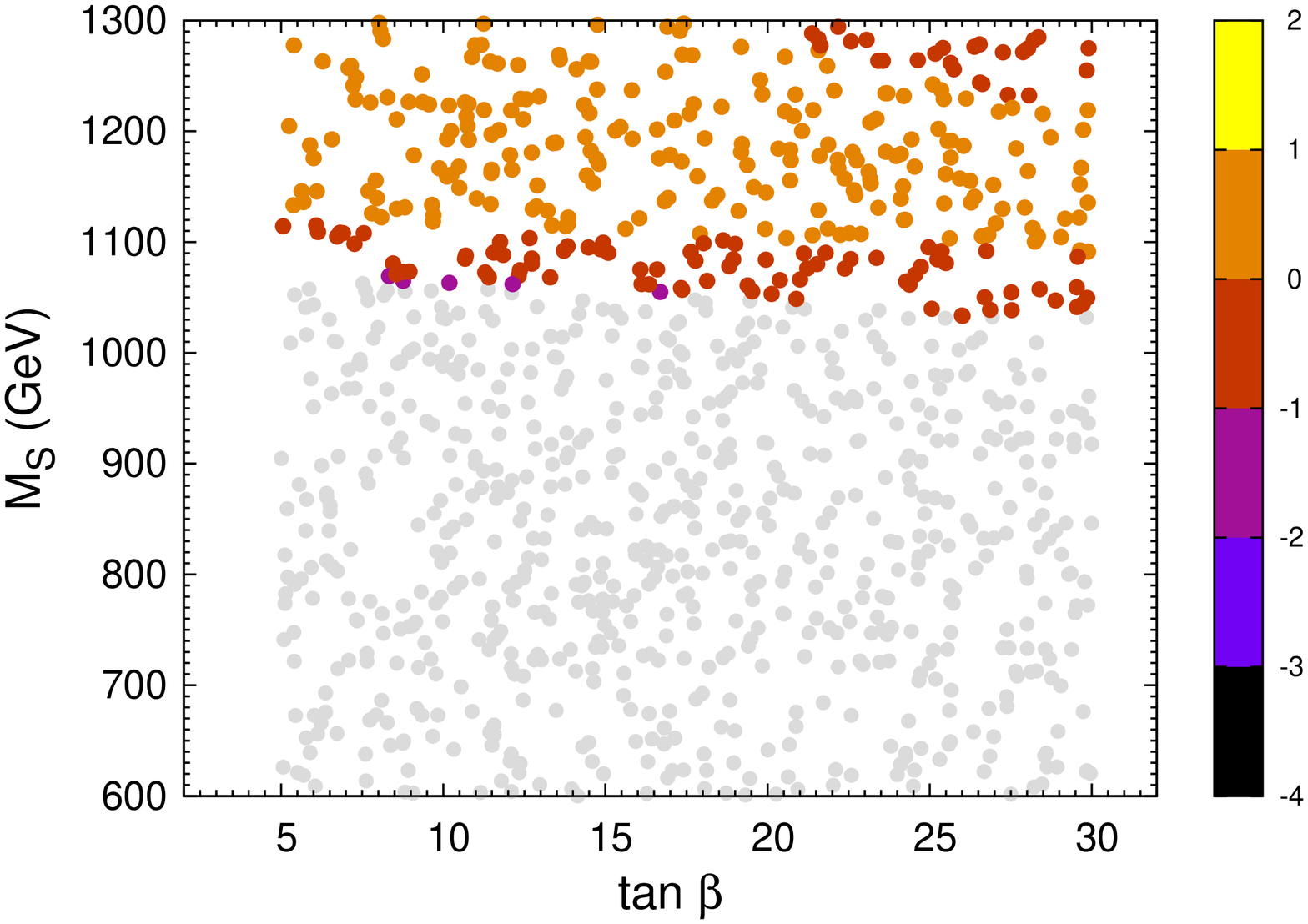,
width=53mm, angle =270}\\
\epsfig{file=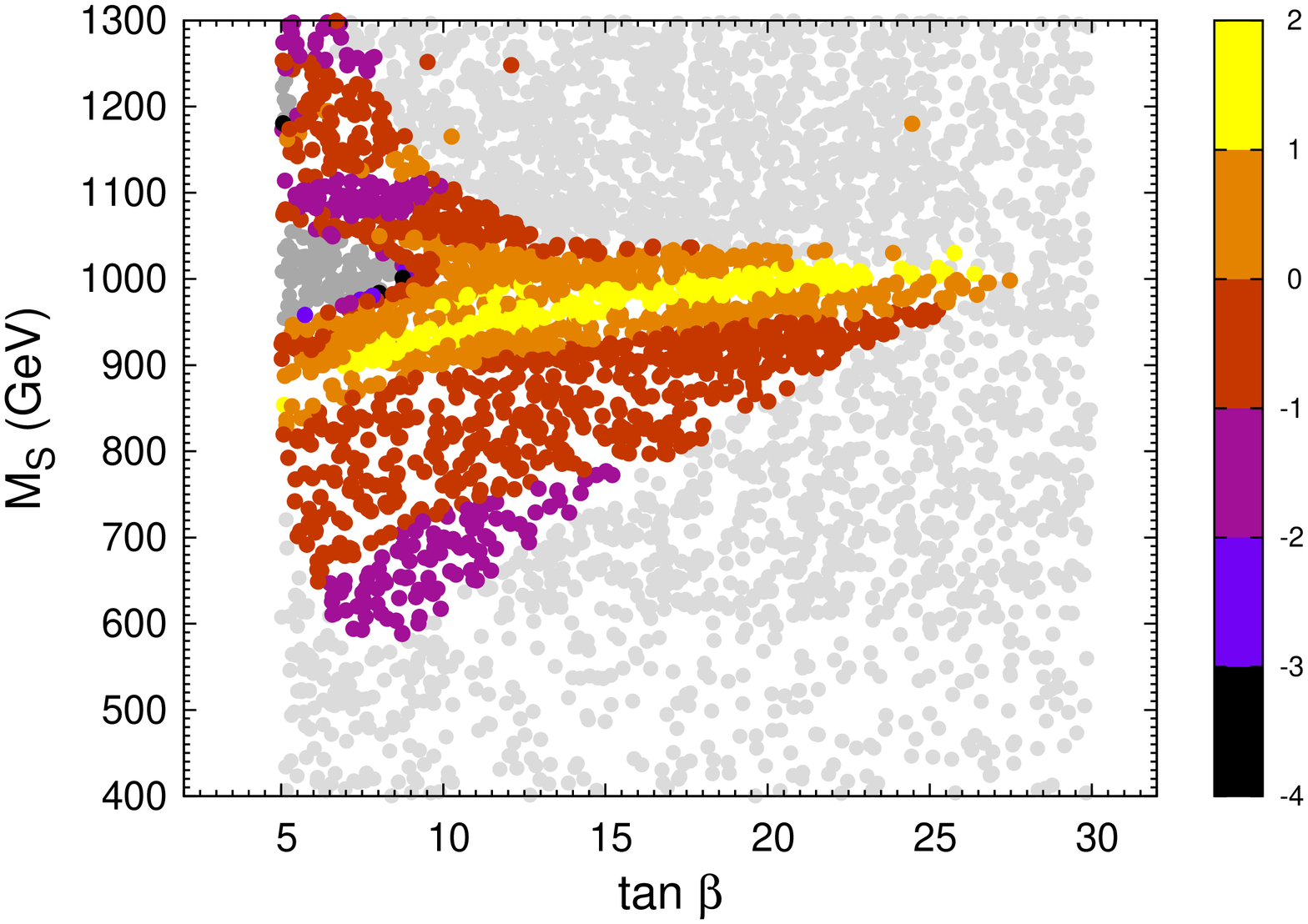,
width=53mm, angle =270} 
\hspace*{2mm}&\hspace*{2mm}
\epsfig{file=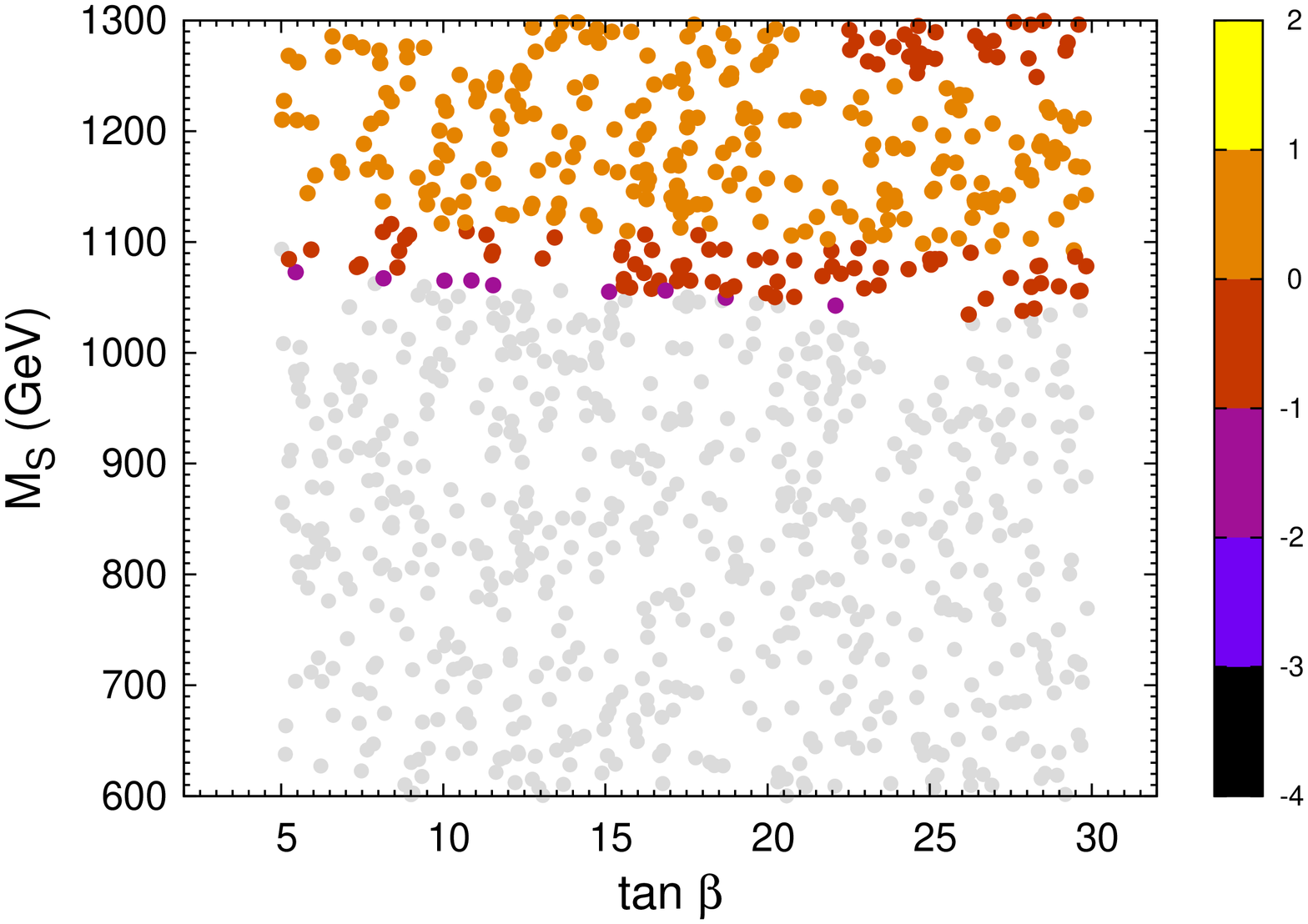,
width=53mm, angle =270}
\end{tabular}
\end{center}
\caption{Relic density in the tan$\beta$ - $M_S$ plane. The colour scheme denotes regimes for the sneutrino LSP relic density, Log$_{10}$ $(\Omega_{\tilde{\nu}_{\rm LSP}} h^2)$. Grey points denote exclusions due to any of the constraints described in Sec.~\ref{sec:num_an}. Dark grey points refer to mass spectra with $m_{\tilde{\nu}_{\rm LSP}}< m_{h^0}/2$. The four panels correspond to the benchmark scenarios P1-P4.}
\label{fig:omega_tanb_MS}
\end{figure}

The panels in Fig.~\ref{fig:omega_tanb_MS} show the dependence of the sneutrino LSP relic density in the tan$\beta$ - $M_S$ plane, for the four benchmark scenarios. Besides the region of parameter space where the sneutrino mass is close to $m_{h^0}/2$, values of the relic density consistent with cosmological observations can be obtained in P3 for $600 \lesssim M_S \lesssim 750$ GeV and $6 \lesssim \rm tan \beta \lesssim 14$ and in P2, P4 for $M_S \sim 1100$ GeV. Notice that the combination of input parameters in scenario P3 allows to get a SUSY breaking scale as low as $M_S \sim 650$ GeV in agreement with the relic density constraint. In contrast, in scenario P1 the solutions with $M_S \lesssim 900$ lead to overabundance of DM, $\Omega_{\tilde{\nu}_{\rm LSP}} h^2 > 0.1$.

\hspace*{-8mm}

\begin{figure}[hbt!]
\begin{center}
\begin{tabular}{cc}
\hspace*{-10mm}
\epsfig{file=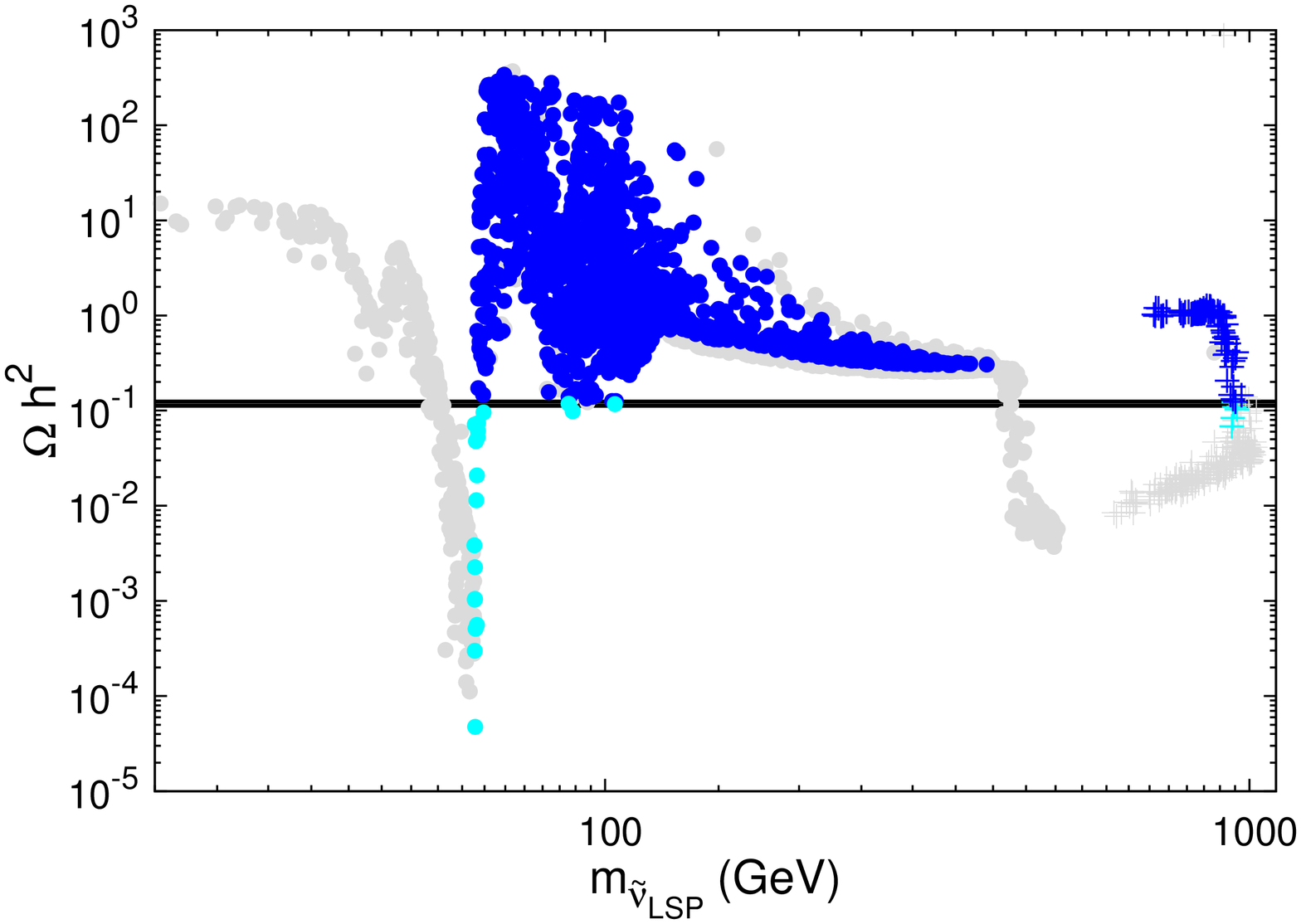,
width=62mm, angle =270} 
&
\epsfig{file=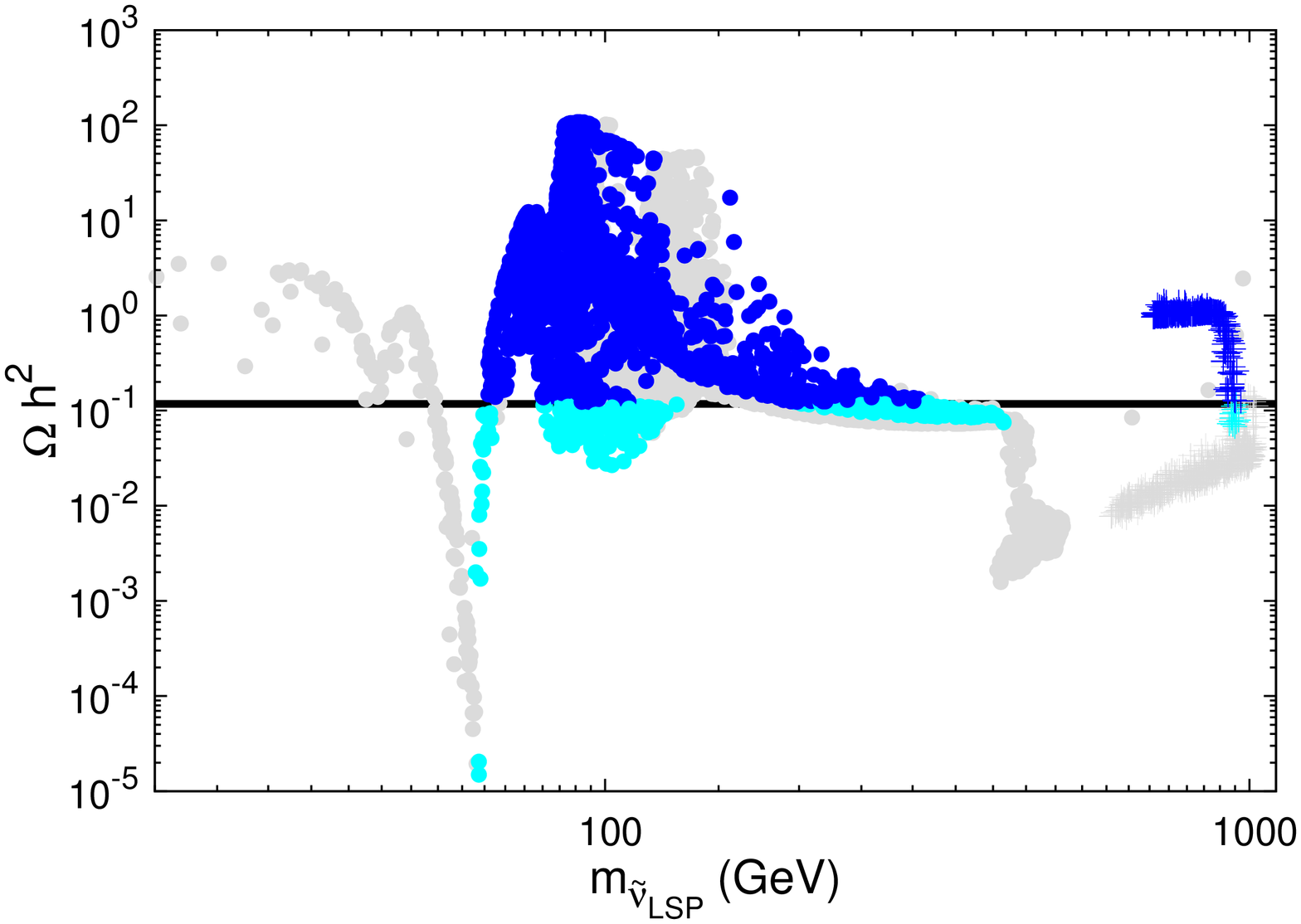,
width=62mm, angle =270}
\end{tabular}
\end{center}
\caption{Sneutrino relic abundance  $\Omega_{\tilde{\nu}_{\rm LSP}} h^2$ as a function of the LSP mixed sneutrino mass $m_{\tilde{\nu}_{\rm LSP}}$. Cyan points lead to viable relic density, whereas the blue ones lead to overabundant DM. Full grey dots are excluded by at least one of the bounds listed in Sec.~\ref{sec:num_an} (including solutions with $m_{\tilde{\nu}_{\rm LSP}}< m_{h^0}/2$).  Scenarios P1 and P2 (P3 and P4) are represented in the left (right) panel: full dots refer to P1 (P3) and crosses to P2 (P4). The black band delimits the $3\sigma$ CL cold DM measurement by the Planck collaboration~\cite{Aghanim:2018eyx}.}.
\label{fig:omega_snumass}
\end{figure}

In Fig.~\ref{fig:omega_snumass} we present values of the sneutrino LSP relic density as a function of its mass. The black lines  in Fig.~\ref{fig:omega_snumass} represent the (thin) $3\sigma$ band of the cold DM density of the Universe as measured by the Planck collaboration~\cite{Aghanim:2018eyx}. Solutions of the numerical scan where the sneutrino LSP has a viable relic density are depicted as cyan points. These either provide the total cold DM in the Universe, or just a fraction. Blue points denote instead solutions which survive the constraints described in Sec.~\ref{sec:num_an}, but that lead to overabundant dark matter. Full dots (crosses) denote scenario P1 (P2) respectively, in the left panel and P3 (P4) in the right one. \\[-.2cm]

Scenario P1, even if characterised by quite large values of $Y_\nu$, in general leads to overabundant sneutrino DM, unless the sneutrino mass is low enough to allow for annihilations via s-channel Higgs exchange. The depletion of the relic density due to this annihilation channel, around $m_{\tilde{\nu}_{\rm LSP}} \sim 60$ GeV, is clearly visible in both panels of Fig.~\ref{fig:omega_snumass}. While efficient annihilations via s-channel Higgs exchange allow for small values of the relic density, when $m_{\tilde{\nu}_{\rm LSP}} \lesssim m_{h^0}/2$ it is very likely for these solutions to be in conflict with current collider limits on BR$(h^0 \to \rm inv)$. Indeed, we have checked that most of these solutions in the left side of the Higgs pole have BR$(h^0 \to \rm inv) \sim 60 \%$ and they are therefore depicted as grey points.  \\[-.2cm]

The annihilation channel via $Z^0$ exchange, although also manifest in both panels of Fig.~\ref{fig:omega_snumass}, is less efficient than the Higgs-mediated one, since the coupling between two scalars ($\tilde{\nu}_{\rm LSP}$) and a vector ($Z^0$) is momentum suppressed. As the sneutrino mass increases, quartic interactions with gauge bosons become effective and, when kinematically allowed, also two-top final states. Hence, for $m_{\tilde{\nu}_{\rm LSP}} \gsim 80$ GeV annihilations into $W^+ W^-$ are particularly important (more than $Z^0 Z^0$) and for  $m_{\tilde{\nu}_{\rm LSP}} \gsim 120$ GeV also the two Higgs final state becomes kinematically accessible. While these annihilations can be quite large, they do not manage to reduce the relic density enough in P1, except for very few solutions at $m_{\tilde{\nu}_{\rm LSP}} \sim 100$ GeV. In contrast, the larger values of $Y_\nu$ characterising P3 allow for solutions in agreement with the Planck constraint around $m_{\tilde{\nu}_{\rm LSP}} \sim 100$ and up to $\sim 500$ GeV. We stress that since there is a strong relation among most parameters of this model -- and in turn among different physical observables --  it is not straightforward to comply with all the constraints applied to the parameter space.  \\[-.2cm]

Scenarios P2 and P4 are characterised by heavier  $m_{\tilde{\nu}_{\rm LSP}}$, which are now degenerate with the rest of the sparticle spectrum. In these cases the mixed sneutrino is not always the LSP (as already noticed in the previous discussion). However, when it is, its relic density would be in general too large unless there are coannihilations, usually with stops. Moreover, and independently of  $m_{\tilde{\nu}_{\rm LSP}}$, since the lightest sneutrinos mass eigenstate is also a CP eigenstate, it coannihilates with the corresponding opposite-CP sneutrino eigenstate (the difference in mass being almost negligible in most of the parameter space considered here)\footnote{This feature provides an interesting realisation of the so called ``inelastic dark matter", that is the DM scattering with nuclei via Z boson exchange can occur inelastically, through a transition between the Re and the Im eigenstates.}.\\

\paragraph{Direct detection}

The sneutrino-nucleus coherent scattering receives two contributions at tree level: the t-channel exchange of a neutral Higgs or of a Z boson (see~\cite{Arina:2008bb,DeRomeri:2012qd} for a detailed description). To compare with experimental results it is convenient to consider the scattering cross section on a single nucleon, multiplied by a factor $\xi = \frac{\Omega_{\tilde{\nu}_{\rm LSP}}}{\Omega_{\rm obs}}$, to rescale the local density of the sneutrino DM to the measured value of cold DM abundance and thus to take into account the possibility that the sneutrino is an underabundant DM candidate. \\[-.2cm]

\begin{figure}[!htb]
\begin{center}
\begin{tabular}{cc}
\epsfig{file=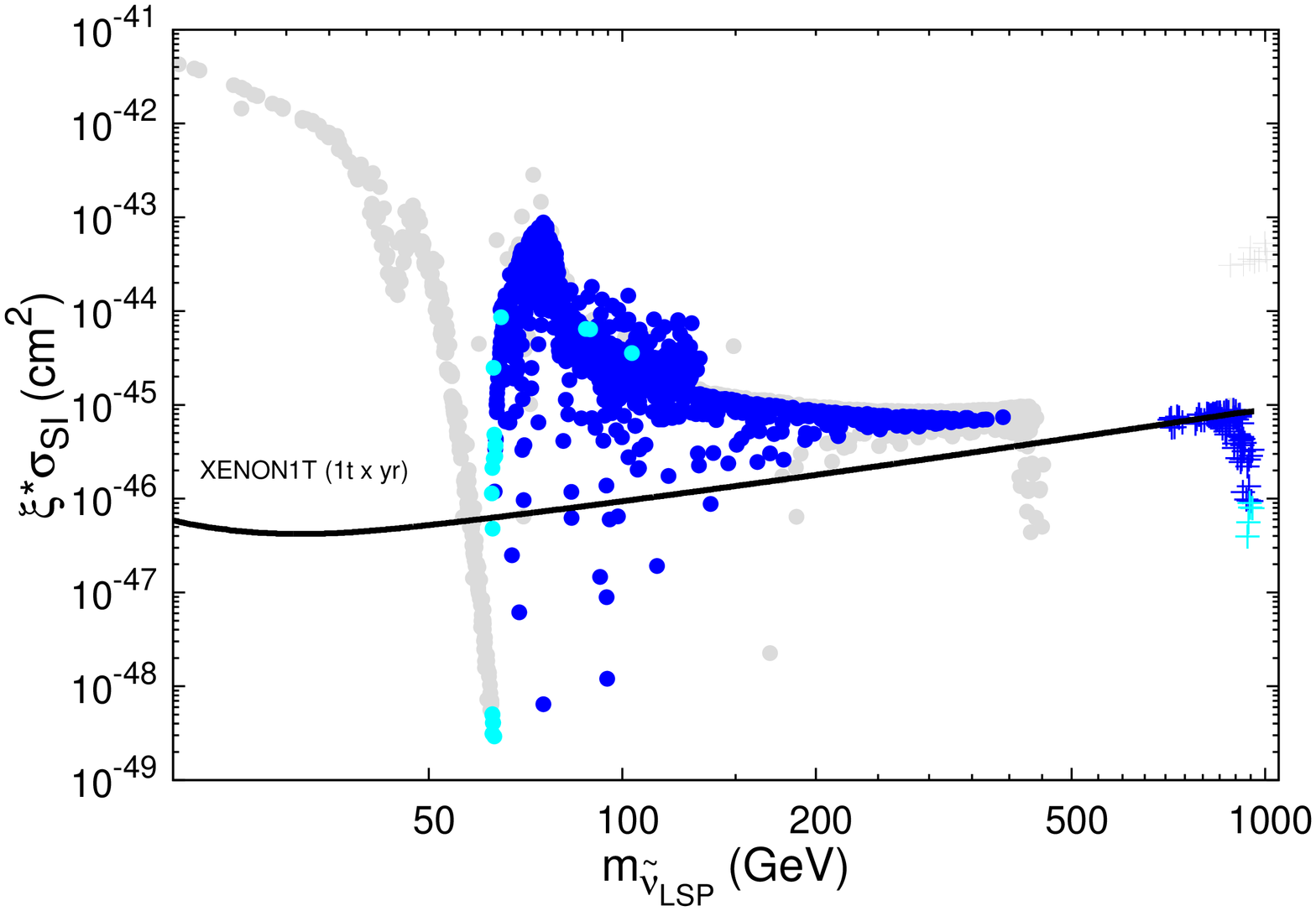,width=58mm, angle =270} 
\epsfig{file=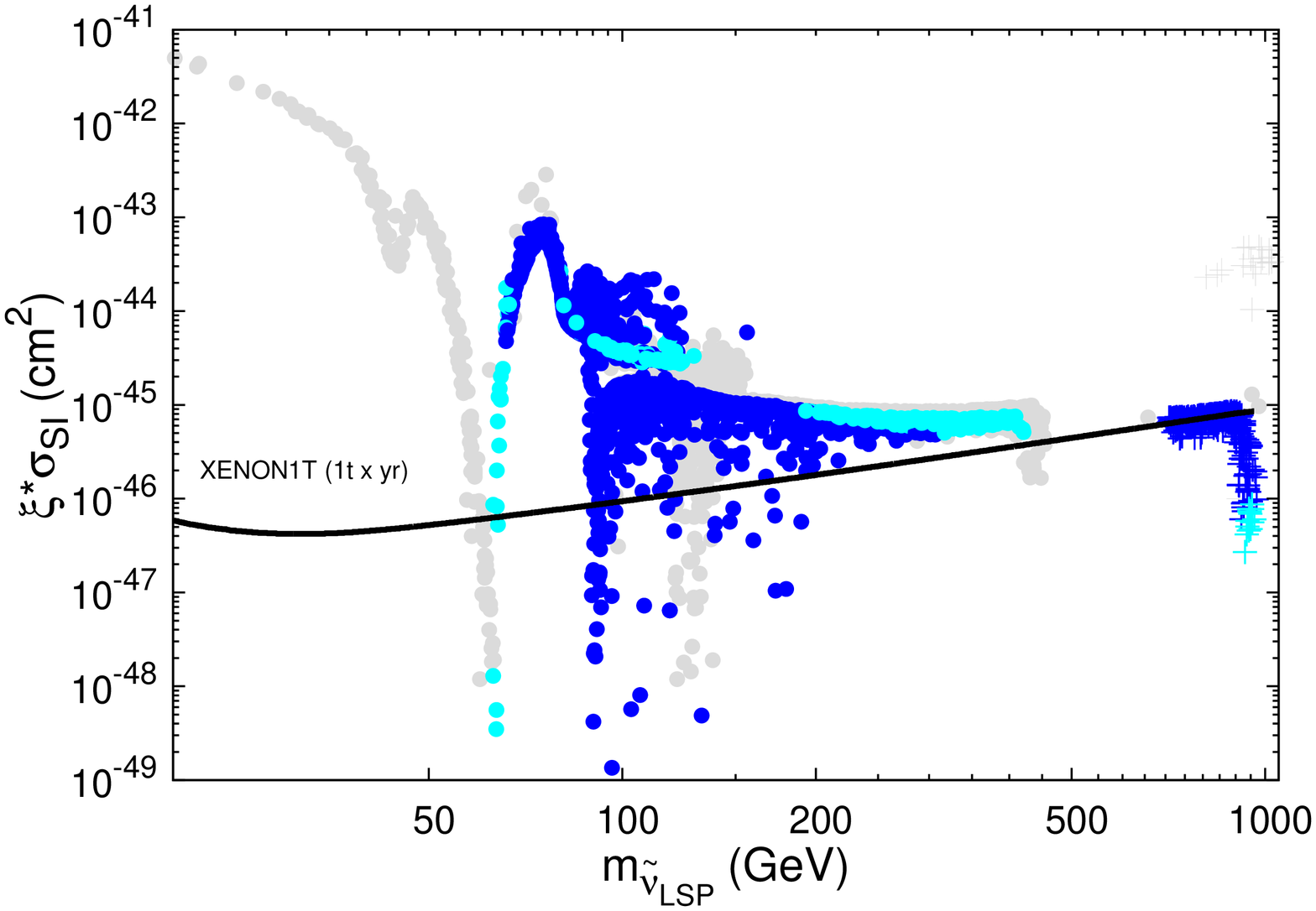,width=58mm, angle =270}
\end{tabular}
\end{center}
\caption{Spin-independent sneutrino-nucleon elastic scattering cross section versus the mixed sneutrino LSP mass. Colours as in Fig.~\ref{fig:omega_snumass}. The plain black line denotes the  most recent bound from XENON1T~\cite{Aprile:2018dbl}. Full grey dots are excluded by at least one of the bounds in Sec.~\ref{sec:num_an} (including solutions with $m_{\tilde{\nu}_{\rm LSP}}< m_{h^0}/2$). Scenarios P1 and P2 (P3 and P4) are represented in the left (right) panel: full dots refer to P1 (P3) and crosses to P2 (P4). }.
\label{fig:sigmaSI_snumass}
\end{figure}

We show in Fig.~\ref{fig:sigmaSI_snumass} the mixed sneutrino spin independent cross section versus the LSP sneutrino mass. The colour code is the same as in Fig.~\ref{fig:omega_snumass}: full dots (crosses) denote scenario P1 (P2) respectively, in the left panel and P3 (P4) in the right one. The cyan points lead to viable relic density, whereas the dark blue lead to overabundant DM. The plain black line denotes the current most stringent limit from XENON1T~\cite{Aprile:2018dbl}. Grey points are excluded by any of the constraints described in Sec.~\ref{sec:num_an}. For a standard isothermal DM halo the current constraint from XENON1T already excludes most of the solutions, except for heavy sneutrinos where coannihilations are important and for the region where the sneutrino annihilates resonantly through the Higgs boson. In the latter case, while unconstrained by the Higgs invisible width (i.e. with $m_{\tilde{\nu}_{\rm LSP}} \gsim m_{h^0}/2$) some allowed solutions lie just below the current XENON1T bound and they could be probed by forthcoming DD data. Moreover, in the near future, DD experiments are planning to push the sensitivity on spin-independent DM-nucleon interaction to the irreducible neutrino background. For instance, an upgrade of XENON1T, XENONnT, should improve upon the current result by more than an order of magnitude, thus probing also the allowed solutions at $m_{\tilde{\nu}_{\rm LSP}} \sim M_S \sim 1$ TeV.

\hspace*{-18mm}

\paragraph{Indirect detection}
Mixed sneutrino DM may also give rise to indirect detection signals. Indeed,
mixed sneutrinos distribute in the galactic halo and they may annihilate in pairs to SM particles, in particular photons, charged leptons and neutrinos. Among the annihilation products which may be searched for, gamma rays are among the most promising messengers, since they preserve the spectral and spatial features of the DM signal. Gamma rays from sneutrino annihilation mostly come from the decay of neutral pions and other mesons produced via hadronization of quarks and gauge bosons. Nevertheless, the gamma-ray signal from sneutrino annihilations is strongly sensitive to the distribution of the DM in the Galaxy.  The computation of the gamma-ray flux therefore depends on assumptions about the DM density profile, which enters quadratically in the integral over the line of sight and the solid angle subtended by the observation (see for instance~\cite{Bertone:2010zza} for a review).

The strongest constraints on the velocity-averaged annihilation cross section $\langle \sigma v \rangle$ and WIMP mass come from the Fermi-LAT analysis of gamma-ray data from dwarf galaxies~\cite{Ackermann:2015zua}. They already exclude annihilation cross sections larger than the expected thermal cross section for DM lighter than $\sim 100$ GeV. At heavier masses, HESS observations towards the center of the Galactic halo impose the most stringent limits~\cite{Abdallah:2016ygi}. 
These constraints are channel dependent, the most stringent bounds being obtained for heavy quarks ($b\bar{b}$) and for the leptonic channel that gives the largest DM gamma-ray flux, $\tau^+ \tau^-$. In our model, the mixed sneutrino annihilates dominantly into $W^+ W^-$, $Z^0 Z^0$, $h^0 h^0$, $b\bar{b}$ and $\tau^+ \tau^-$, with different branching ratios and dependent on the mass regime. 
While a thorough analysis of indirect detection signals from mixed sneutrino annihilations is out of the scope of this work, we have checked that the solutions which survive all the other constraints (see Sec.~\ref{sec:num_an}) in general lead to $\xi^2 \langle \sigma v \rangle$ (also weighted by the respective branching fraction) which lie below the current limits from Fermi-LAT~\cite{Ackermann:2015zua}.
While at present the constraints from direct detection experiments seem to be more important, future high-energy low-threshold gamma-ray space experiments such as eASTROGAM~\cite{DeAngelis:2016slk} or COMPAIR~\cite{Moiseev:2015lva} together with ground-based telescopes like the High-Altitude Water Cherenkov Observatory (HAWC)~\cite{Abeysekara:2014ffg}, the Large High Altitude Air Shower Observatory (LHAASO)~\cite{DiSciascio:2016rgi} and the Cherenkov Telescope Array (CTA)~\cite{Acharya:2013sxa} will further probe the mixed sneutrino DM scenario.

\section{Conclusions and outlook}
\label{sec:conclusions}

We have considered supersymmetric version of the inverse seesaw mechanism in the context of a low energy effective theory, where soft SUSY breaking arises from a higher dimensional theory in which the extra spatial dimension is compactified on an orbifold. If the compactification scale lies close to the SUSY breaking scale, a compressed SUSY spectrum with approximate degeneracy amongst sparticle masses can arise. We have built upon  previous analyses -- performed within the simplest Degenerate MSSM with massless neutrinos -- by implementing the inverse seesaw mechanism, in order to accommodate neutrino masses and mixings as required by the current oscillation data. Other low-scale seesaw scenarios, such as the linear seesaw mechanism~\cite{Akhmedov:1995ip,Akhmedov:1995vm,Malinsky:2005bi}, can also be envisaged and analysed in a similar manner.

By means of detailed numerical analyses, we have shown that the implementation of the inverse seesaw mechanism within a compact SUSY context leads to important effects. First of all, the large neutrino Yukawa couplings introduced in this framework allow us to accommodate the observed value of the Higgs boson mass in a more natural way, compared to the simplest weak-scale SUSY scenario without neutrino masses. Second, the effective SUSY breaking scale can be lowered thanks to the combined effect of having a compressed super-particle spectrum, as well as the presence of the new Yukawa couplings characterizing neutrino mass generation in the inverse seesaw mechanism. We have found that the scale characterizing the compressed supersymmetric spectrum can be as low as 500-600 GeV for $\tan\beta\sim 10$, even after taking into account the most relevant experimental restrictions from collider, high-intensity and cosmological observations. Last but not least, the inverse seesaw mechanism also implies that a mixture of isodoublet and isosinglet sneutrinos can be the LSP, thus allowing for a novel WIMP dark matter candidate, besides the standard neutralino. Interestingly, in some regions of the parameters the sneutrino DM can be quite light, with mass $\lesssim 100$ GeV, while the remaining SUSY spectrum is approximately degenerate.

Besides the searches for supersymmetric partners, we note that inverse seesaw schemes with sneutrino-like dark matter have their own collider implications. Indeed, these low-scale seesaw schemes offer the tantalizing possibility of searching directly for the messengers of neutrino mass generation at collider energies. This task has been taken up since the LEP days~\cite{Dittmar:1989yg,Abreu:1997pa}. Dedicated searches for the quasi-Dirac heavy neutrinos typical of these schemes can be conducted using proton-proton collisions at the LHC, see for instance \cite{Das:2017zjc,Dib:2017vux}. The results of a recent search were reported in~\cite{Sirunyan:2018mtv}. Prospects for probing these heavy neutrinos at future experiments such as SHiP, FCC-ee or CEPC have also been discussed, see~\cite{Drewes:2015vma} and references therein. \\[-.3cm]

In summary, we have considered a well-motivated theoretical framework which addresses open problems concerning the origin of neutrino masses and the naturalness of the Higgs boson mass. In addition to the theory motivations, this model leads to interesting phenomenological features. The low-scale SUSY spectrum will be further probed at the LHC through direct searches. However, we stress that dedicated collider searches should take into account specific features of this scenario, both spectrum and couplings, associated to the presence of the inverse seesaw mechanism. They imply that the lightest supersymmetric particle is expected to be a mixture of isodoublet and isosinglet sneutrinos. In addition to such specific collider implications, our sneutrino-like dark matter scenario will be further probed by dedicated direct as well as indirect search experiments.

\section*{Acknowledgements}

Work funded by the Spanish grants FPA2017-85216-P (AEI/FEDER, UE), SEV-2014-0398, FPA2017-90566-REDC and PROMETEOII/2014/084 (Generalitat Valenciana). V.D.R. is grateful to Alexander Pukhov for valuable help with Micromegas. V.D.R. acknowledges financial support from the ``Juan de la Cierva Incorporaci\'on" program (IJCI-2016-27736). The work of K.M.P. was partially supported by SERB Early Career Research Award (ECR/2017/000353) and by a research grant under INSPIRE Faculty Award (DST/INSPIRE /04/2015/000508) from the Department of Science and Technology, Government of India. K.M.P. thanks AHEP group at IFIC, Valencia for hospitality where this work was started. He also thanks the DESY and CERN theory groups where part of this work was carried out.

\end{document}